\documentclass[pdflatex,sn-mathphys-num]{sn-jnl}

\usepackage{graphicx}%
\usepackage{multirow}%
\usepackage{amsmath,amssymb,amsfonts}%
\usepackage{amsthm}%
\usepackage{mathrsfs}%
\usepackage[title]{appendix}%
\usepackage{xcolor}%
\usepackage{textcomp}%
\usepackage{manyfoot}%
\usepackage{booktabs}%
\usepackage{algorithm}%
\usepackage{algorithmicx}%
\usepackage{algpseudocode}%
\usepackage{listings}%
\usepackage{tabularx}%
\usepackage{float}
\usepackage{subcaption}

\lstdefinestyle{jsonstyle}{
    backgroundcolor=\color{black!5},
    basicstyle=\ttfamily\scriptsize,
    breaklines=true,
    frame=single,
    rulecolor=\color{black!40},
    string=[s]{"}{"},
    stringstyle=\color{blue},
    comment=[l]{:},
    commentstyle=\color{black},
    morestring=[b]',
}

\lstdefinestyle{datastyle}{
    backgroundcolor=\color{black!5},
    basicstyle=\ttfamily\scriptsize,
    breaklines=true,
    frame=single,
    rulecolor=\color{black!40},
}

\definecolor{cypher-kw}{RGB}{0,0,180}
\definecolor{cypher-rel}{RGB}{160,0,0}
\definecolor{cypher-str}{RGB}{0,120,0}
\definecolor{cypher-bg}{RGB}{248,248,248}

\lstdefinelanguage{Cypher}{
  keywords={MATCH,OPTIONAL,WITH,WHERE,RETURN,ORDER,BY,LIMIT,AS,DESC,ASC,
            CASE,WHEN,THEN,ELSE,END,NOT,IN,AND,OR,UNWIND,COLLECT,COUNT,
            DISTINCT,SUM,AVG,MIN,MAX,CALL,UNION,ALL,IS,NULL},
  sensitive=true,
  morestring=[b]',
  morestring=[b]",
  morecomment=[l]{//},
}

\definecolor{darkblue}{rgb}{0,0,0.55}

\IfFileExists{tcolorbox.sty}{%
  \usepackage{tcolorbox}
  \newtcolorbox{researchbox}{
    colback=blue!5!white,
    colframe=blue!75!black,
    fonttitle=\bfseries,
    title=Research Question,
    width=\linewidth,
    arc=2mm,
    boxrule=0.5pt
  }
}{%
  \newenvironment{researchbox}{%
    \par\medskip\noindent\begin{center}\begin{minipage}{0.97\linewidth}%
    \rule{\linewidth}{0.5pt}\par\textbf{Research Question}\par\smallskip}{%
    \par\rule{\linewidth}{0.5pt}\end{minipage}\end{center}\medskip}
}

\begin{document}

\title[Quantum Patterns in Practice]{
From Pattern Detection to Composition Analysis in Quantum Software
}

\author*[1]{\fnm{Neilson C. L.} \sur{Ramalho}}\email{neilson@usp.br}

\author[1]{\fnm{Erico A.} \sur{da Silva}}\email{augusto.ericosilva@usp.br}

\author[3]{\fnm{Anthony} \sur{Accioly}}\email{a.accioly@alumni.usp.br}

\author[2]{\fnm{Higor Amario} \sur{de Souza}}\email{higoramario@usp.br}

\author[1]{\fnm{Marcos Lordello} \sur{Chaim}}\email{chaim@usp.br}

\affil*[1]{\orgdiv{School of Arts, Sciences, and Humanities}, \orgname{University of S\~{a}o Paulo}, \orgaddress{\city{S\~{a}o Paulo}, \state{SP}, \country{Brazil}}}

\affil[2]{\orgdiv{Department of Computer and Digital Systems Engineering, Polytechnic School}, \orgname{University of S\~{a}o Paulo}, \orgaddress{\city{S\~{a}o Paulo}, \state{SP}, \country{Brazil}}}

\affil[3]{\orgdiv{Alumni}, \orgname{University of S\~{a}o Paulo}, \orgaddress{\city{S\~{a}o Paulo}, \state{SP}, \country{Brazil}}}

\abstract{

Quantum software patterns provide high-level abstractions for building quantum programs, but there is still little empirical evidence on how they are adopted in practice. In prior work, we extended an existing quantum-pattern atlas into a 61-pattern catalog, created a knowledge base that links framework components to those patterns, and built a tool that mines pattern implementations from open-source code. We applied this tool on 80 projects and find that all 23 patterns occur in practice. In this work, we extend the tool with two additional matching channels and a vocabulary expansion step, and execute a quantitative evaluation of its accuracy on Qrisp, a framework not present in the knowledge base, reaching a micro-F1 of 0.712 against 0.449 without the expansion step. We then construct composition graphs that record calls among the high-level framework components associated with patterns and store them in a graph database. We use these graphs to examine how pattern implementations are assembled inside each framework, why patterns co-occur, and how much of a pattern's detection count comes from components called directly by developers rather than introduced through internal framework calls. We release qpa, an open-source mining pipeline, together with the knowledge base, which maps 286 framework components across five sources to the pattern catalog, maintained with the support of an LLM ensemble that classifies newly extracted components, and the resulting pattern usage dataset, to support reproducible studies on the adoption and evolution of quantum patterns.

}


\maketitle

\section{Introduction}

Quantum computing (QC) has developed from a largely theoretical field into an active area of research and software development. Advances in hardware and techniques for reducing noise effects have encouraged work on applications in areas such as molecular simulation, cybersecurity, and finance. Developing quantum software nevertheless remains difficult. Quantum programs are often expressed in terms of circuits, and understanding their behavior requires familiarity with quantum-mechanical principles such as superposition, entanglement, interference, and probabilistic measurement~\cite{Hidary2019Quantum}. The combination of quantum-mechanical knowledge and software development skills makes quantum software development an interdisciplinary task and creates challenges for software engineers entering the field \cite{WhenSEMeetsQC}.

Classical software engineering has developed tools, processes, and abstractions for managing the complexity of software development. Among these abstractions, software patterns document reusable solutions to recurring design problems~\citep{DesignPatternsGof}. Comparable practices and abstractions for quantum software are still developing, and establishing them is one of the concerns of Quantum Software Engineering (QSE)~\citep{MurilloQSE2025}. One part of this effort is the definition of quantum computing patterns~\citep{LeymannQuantumAlgorithms,PatternsCircuitCutting2023,HarnessingPatterns2025,PatternsQuantumErrorHandling2022,Georg2023_PatternsQuantumExecution,Buehler2023_QuantumSoftwareEngineeringPatterns,Stiliadou2025_QMLPatterns}. These patterns describe reusable solutions at different levels of abstraction, from circuit construction and execution to the organization of hybrid quantum-classical applications. Although a growing number of patterns have been documented, there is still limited empirical evidence about their use in practitioner code. Studying their adoption presents two practical challenges: mapping abstract pattern definitions to concrete implementations in quantum frameworks and detecting those implementations at scale in open-source projects. This paper goes beyond identifying quantum patterns in practitioner code. Its main contribution is a graph-based representation of calls among the framework components associated with these patterns. This representation shows how pattern implementations are assembled inside each framework.

A recent study took a first step toward closing this gap~\citep{ramalho2026mining}. It extended the PlanQK catalog~\citep{PlanQK_QuantumPatterns_2024}, a structured collection of quantum-pattern definitions, and used it to build a knowledge base (KB) from Qiskit~\citep{qiskit_community}, PennyLane~\citep{PennyLanePaper}, and Classiq~\citep{classiq-library2024}. The KB maps concrete framework components to abstract patterns. Here, a framework component is a function, method, or class that implements a reusable quantum-related routine. Each KB entry contains the component's name, documentation, source framework, and assigned pattern. The extended catalog contains 61 patterns, but a pattern only enters the KB when at least one framework component implements it, according to the manual classification of framework components (Section~\ref{sub:pattern_kb}). Of the 61 patterns, 23 meet this condition, so the detection and composition analyses reported here cover that subset rather than the full catalog.

The study also introduced qpa (Quantum Patterns Analyzer), an open-source pipeline that discovers quantum projects on GitHub, converts their notebooks into Python scripts, and searches the resulting code for evidence of the components stored in the knowledge base. The earlier version used two matching channels, with each channel representing a different type of evidence. The name channel compared function calls in practitioner code with component names in the KB, while the documentation channel compared notebook text with component documentation using semantic similarity. Each successful match was translated into the pattern assigned to that component. The resulting usage dataset recorded these detections and was used to measure how frequently patterns appeared in open-source notebooks.

That study showed that the patterns occur in practice, but it considered only detection counts: it did not measure how accurate those counts were, and it treated each detection in isolation, without examining how the detected components relate to one another inside a framework. The present study addresses both limitations by evaluating detection accuracy and analyzing calls among framework components.

Building on the catalog, KB, and pipeline from the previous study, we construct composition graphs that record calls among the high-level framework components associated with patterns and store them in a graph database. These call relationships show how pattern implementations are assembled inside each framework. They allow us to distinguish patterns called directly by developers from patterns introduced through internal framework calls, explain why patterns co-occur, and show whether a framework exposes a pattern as a single high-level component or as a sequence of smaller components.

We also improve the detection analysis. To keep the KB current as frameworks evolve, we add a semi-automated classification step in which three LLMs from different providers vote on the pattern label of each newly extracted component. We extend the KB from three framework sources to five by adding Qiskit Algorithms~\citep{qiskit2024} and Qiskit Machine Learning~\citep{QiskitML}. We increase the number of matching channels from two to four and introduce a two-phase vocabulary-expansion step that allows the tool to recognize components whose names differ from those in the KB. These improvements enable the first quantitative evaluation of the tool on Qrisp~\citep{qrisp}, an open-source QC framework that provides high-level programming abstractions based on typed quantum variables and automatic uncomputation. On this evaluation, vocabulary expansion raises micro-F1 from 0.449 to 0.712, with a micro-precision of 0.913.

These two parts of the study address limitations of the previous work and lead to three research questions:

    \begin{itemize}
        \item [\textbf{RQ1.}] How accurately does the improved tool detect quantum patterns, and does the vocabulary expansion step help when the target framework's naming differs from the knowledge-base frameworks?
        \item [\textbf{RQ2.}] Do patterns detected in a file reflect deliberate choices by the developer, or are some introduced by the framework's internal composition?
        \item [\textbf{RQ3.}] Does each framework expose the same pattern as a single high-level routine or as a chain of smaller components?
    \end{itemize}

 Our work makes the following contributions to the QSE research community:

    \begin{itemize}

        \item \textbf{Composition graphs for analyzing pattern implementations:} a graph-based representation of calls among the high-level framework components associated with quantum patterns. Analyzing these relationships allows us to distinguish direct developer calls from components introduced through internal framework calls, explain why patterns co-occur, and compare how the five frameworks expose the same pattern.

        \item \textbf{Updated empirical evidence on pattern adoption:} a practitioner-focused analysis based on 611 pattern detections across 358 files from 904 Python scripts extracted from Jupyter Notebooks in a selected set of 80 open-source projects after excluding three framework-maintained example repositories. The analysis shows that all 23 catalog patterns represented in the KB occur in practice and that usage concentrates on utility and variational patterns across three levels of abstraction.

        \item \textbf{Semi-automated KB classification:} an LLM ensemble that classifies newly extracted framework components into the 61-pattern catalog. Three models from different providers vote on each component and components without a majority label are left for manual review. In this study, the ensemble classified the 50 components added to the KB since the previous manual categorization. 
        
        \item \textbf{Supporting artifacts:} the qpa tool together with the datasets and graphs that enable these analyses, packaged for reproducible reuse. We release (i) a knowledge base linking 286 components from five quantum framework sources (Qiskit, PennyLane, Classiq, Qiskit Algorithms, and Qiskit Machine Learning) to the pattern catalog; (ii) the pattern usage dataset; and (iii) the composition graphs and the pipeline that builds them. The pipeline ships as a Docker Compose~\cite{DockerCompose} setup that provisions the graph database and reconstructs the graphs from fixed framework versions, so a reader can either query the released graphs directly or rebuild them from source. All artifacts are available in the qpa replication package\footnote{qpa replication package: \url{https://github.com/saeg/qpa-v2}.}.
    \end{itemize}

    The new analysis shows that raw detection counts do not always represent patterns selected directly by developers. Some patterns are introduced by calls made inside a framework, and frameworks expose the same pattern at different levels of granularity. The expanded KB also detects additional instances of variational and quantum machine learning patterns that were outside the scope of the previous KB. Alongside these new findings, the results corroborate the previous study by showing that developers use patterns at three levels of abstraction, from low-level utilities to domain-specific applications. The composition graphs, tool, and datasets support further studies of how quantum patterns are implemented and adopted and how their framework components are combined.

    The rest of the paper is organized as follows. Section~\ref{sec:background_related_work} discusses the background and related work on quantum software patterns. Section~\ref{sec:tool} describes the pattern knowledge base, the selection of projects and notebooks, the matching process, the patterns found in open-source projects, and the evaluation of the pattern detector. Section~\ref{sec:composition} introduces the composition graphs and examines what they reveal about direct and indirect pattern use, pattern co-occurrence, and differences in how frameworks expose the same pattern. Section~\ref{sec:usage} discusses threats to validity, and Section~\ref{sec:conclusion} concludes the paper.

    \section{Background and related work}
    \label{sec:background_related_work}

    A quantum computer uses properties of quantum mechanics, such as superposition, entanglement, and interference, to perform computations. For some problems, quantum algorithms may offer advantages over known classical approaches~\citep{Hidary2019Quantum}. However, the intersection of these quantum principles with computation introduces challenges to classical software engineering practices. While classical software engineering has matured over decades, practices for Quantum Computing (QC) are still emerging. In response, the research community in Quantum Software Engineering is actively creating higher-level abstractions and defining reusable solutions, known as patterns \citep{LeymannQuantumAlgorithms, PatternsCircuitCutting2023, HarnessingPatterns2025, PatternsQuantumErrorHandling2022, Georg2023_PatternsQuantumExecution, Buehler2023_QuantumSoftwareEngineeringPatterns, Stiliadou2025_QMLPatterns}, to help practitioners develop complex hybrid quantum programs.

    The effort to catalog these solutions began with foundational work by \citet{LeymannQuantumAlgorithms}, who defined an initial set of core patterns for quantum algorithm development. This initial set includes building blocks such as Initialization, Uniform Superposition, Creating Entanglement, Oracle, Amplitude Amplification, and Uncompute. This initial body of patterns has since been extended to address practical challenges. For example, patterns for error management have been introduced, including Error Correction, Readout Error Mitigation, and Gate Error Mitigation \citep{PatternsQuantumErrorHandling2022}. Other work has focused on hybrid solutions and reusability, introducing patterns like Quantum Module, Hybrid Module, and Quantum Circuit Translation to better structure and integrate quantum and classical code \citep{Buehler2023_QuantumSoftwareEngineeringPatterns}.

    Research on quantum patterns has also expanded into specific application domains. In Quantum Machine Learning, \citet{Stiliadou2025_QMLPatterns} document patterns such as Quantum Clustering, Quantum Classification, and Quantum Neural Network. Similarly, in Quantum Artificial Intelligence, \citet{klymenko2024architecturalpatternsdesigningquantum} identify architectural patterns for designing hybrid systems, grouped into categories like Quantum-Classical Split and Quantum Middleware Layer. At a higher level of abstraction, architectural patterns for entire quantum software systems have also been proposed by \citet{aktar2025decisionmodelsselectingarchitecture}, focusing on concerns like Communication, Decomposition, Fault Tolerance, and Algorithm Implementation.
    
    Quantum patterns are also collected in resources such as the Quantum Computing Patterns Atlas, which documents 59 patterns~\citep{PlanQK_QuantumPatterns_2024}. Following the approach defined by \citet{DesignPatternsGof}, each pattern is cataloged with its intent, context, and solution. The atlas covers multiple levels of abstraction, from basic circuit-level operations like Circuit Cutting to reusable algorithmic blocks like Quantum Fourier Transformation.

    Empirical studies have also investigated how quantum patterns appear in source code. \citet{fernandez2025exploring} conducted a case study of 2,610 Qiskit programs, focusing on four foundational patterns: Initialization, Superposition, Entanglement, and Oracle. Similarly, \citet{CastilloPreliminaryStudy2024} analyzed 80 source files written in Qiskit and OpenQASM and found that Initialization and Uniform Superposition were the most frequently implemented patterns. \citet{shen2025quantumpatterndetectionaccurate} developed a tool that combines static and dynamic analysis to detect eight patterns, including Basis Encoding, Angle Encoding, Amplitude Encoding, and Quantum Phase Estimation, in benchmark programs from sources such as MQT Bench~\citep{Quetschlich_2023}.
    
    Most closely related to the present work, a prior empirical study~\citep{ramalho2026mining} analyzed 985 Jupyter Notebooks from 80 open-source projects using an earlier version of qpa with three framework sources and two matching channels. The present paper reuses the pattern catalog, KB structure, and basic detection process from that study. It extends the KB with two additional framework sources, adds two matching channels and vocabulary expansion, and evaluates the detector against a manually labeled dataset. Its main contribution is a set of composition graphs representing calls among framework components associated with patterns. Section~\ref{sec:tool} describes the detection process and the patterns found in open-source projects, while Section~\ref{sec:composition} presents the composition graphs and the corresponding analysis.

\section{Pattern detection in open-source projects}
\label{sec:tool}

This section describes the pattern detections used as input to the composition analysis in Section~\ref{sec:composition} and evaluates their accuracy for RQ1. The qpa tool and its original mining workflow were introduced in previous work~\citep{ramalho2026mining}, so we summarize the published workflow and focus on the extensions made for this study: an updated list of projects, an expanded five-source KB, two additional matching channels, and vocabulary expansion. We then evaluate the extended detector on Qrisp and report the updated results that contextualize the composition analysis.

At a high level, qpa connects three types of information. The pattern catalog provides the abstract pattern definitions. Selected quantum frameworks provide concrete API components, such as functions, methods and classes, which are assigned to these patterns and included in the seed KB. Practitioner notebooks are then converted into Python scripts and compared with the KB. Evidence found in a script is associated with the corresponding pattern and recorded as a pattern detection. In this paper, we keep this basic data flow, but extend the framework sources and matching procedure.

\subsection{The pattern knowledge base}
\label{sub:pattern_kb}

In this work, we use the 61-pattern catalog introduced in the previous study~\citep{ramalho2026mining}. This catalog was derived from 59 pattern definitions collected through the PlanQK Pattern Atlas API~\citep{PlanQK_QuantumPatterns_2024,PatternAtlasGitHub}\footnote{The PlanQK site that hosted these definitions has since been taken down, so we republished the 59 pattern definitions as a static reference list at \url{https://qpa-quantum-patterns.web.app/}. The Pattern Atlas software remains open source, but no public instance of the API is currently online.} and was subsequently refined by consolidating related patterns and adding patterns identified through framework analysis, as described in the previous study. The original 59 Atlas records are stored in the replication package as the JSON file \texttt{data/quantum\_patterns.json}\footnote{\url{https://github.com/saeg/qpa-v2/blob/main/data/quantum_patterns.json}.}, whereas the extended catalog provides the pattern labels used to classify framework components in the KB. Of its 61 patterns, 23 are represented by at least one component in the current KB and therefore fall within the scope of the detection and composition analyses.

    \subsubsection{Framework components}
    The seed KB contains components collected from five framework sources. In this work, we kept Qiskit~\citep{qiskit2024}, PennyLane~\citep{PennyLanePaper}, and Classiq~\citep{classiq-library2024}, which were selected and analyzed in the previous study~\citep{ramalho2026mining}. We then added Qiskit Algorithms~\citep{qiskit_community} and Qiskit Machine Learning~\citep{QiskitML} to extend the KB with algorithm-level components and quantum machine learning operations.

    The selection was based on a practical criterion: each source organizes documented, reusable API components in identifiable packages, modules, or public interfaces, allowing qpa to extract them automatically. Together, these sources provide circuit-library components, reusable templates, high-level synthesis functions, quantum algorithm implementations, and quantum machine learning routines. Cirq~\citep{cirq_zenodo_2025} and TensorFlow Quantum~\citep{broughton2021tensorflowquantumsoftwareframework} were not included because their code organization would require a different extraction procedure.

    For Classiq, PennyLane, and Qiskit, we followed the extraction procedures defined in the previous study~\citep{ramalho2026mining}. The Classiq extractor reads the documented components exposed through the public interface of \texttt{classiq.open\_library.functions}. The PennyLane extractor collects documented classes and functions from \texttt{pennylane.templates}, while the Qiskit extractor collects them from \texttt{qiskit.circuit.library} after excluding standard gates and deprecated or redundant entries. Rerunning these extractors produced 64 Classiq components, 68 PennyLane components, and 85 Qiskit components in the current KB.

    From \texttt{qiskit-algorithms}, qpa scans the packages for amplitude amplification, amplitude estimation, eigensolvers, minimum eigensolvers, phase estimation, state fidelity, and time evolution. It extracts documented, non-private classes while excluding classical eigensolvers and optimizers, numerical solvers, and classes used as result, error, job, or other support containers. Applied to the pinned package version, this procedure produced 39 components. These include algorithm implementations such as \texttt{VQE}, \texttt{QAOA}, \texttt{Grover}, and \texttt{PhaseEstimation}, as well as interfaces and supporting classes used with these algorithms. Their inclusion allows qpa to detect direct uses of named algorithm APIs that were not covered by the previous KB.

    For \texttt{qiskit-machine-learning}, qpa reads the public exports declared through \texttt{\_\_all\_\_} in ten selected namespaces covering neural networks, classifiers and regressors, quantum kernels, circuit construction, connectors, gradients, quantum inference, and state fidelity. It follows each exported name to its source definition and extracts its documentation. Exports beginning with \texttt{Base}, exports ending in \texttt{Result}, \texttt{Error}, \texttt{Job}, \texttt{Type}, \texttt{Factory}, \texttt{Protocol}, or \texttt{Mixin}, and the support interface \texttt{TrainableKernel} are excluded. Applied to the pinned package version, this procedure produced 30 components, including \texttt{EstimatorQNN}, \texttt{SamplerQNN}, \texttt{VQC}, \texttt{QSVC}, and \texttt{FidelityStatevectorKernel}. These components represent Data Encoding, Domain-Specific Application, Quantum Neural Network, SWAP Test, and Variational Quantum Algorithm.

    For both added sources, qpa reruns these selection procedures and compares the extracted component names with the curated KB and its component data files. The check reports new, removed, renamed, duplicated, or unassigned components, as well as new API namespaces that require a scope decision. For the pinned versions used in this study, it reproduced the sets of 39 Qiskit Algorithms components and 30 Qiskit Machine Learning components exactly.

    \subsubsection{Assigning components to patterns}
    Each framework component in the KB is assigned to exactly one pattern. For Qiskit, PennyLane, and Classiq, we reuse the classifications produced in the previous study~\citep{ramalho2026mining}. In that study, three raters independently assigned a pattern to each component and resolved disagreements through discussion. The reported agreement was Fleiss'~$\kappa = 0.8171$ and Light's~$\kappa = 0.8173$. Pairwise Cohen's~$\kappa$ values were 0.7261 for Raters~1 and~2, 0.9223 for Raters~1 and~3, and 0.803 for Raters~2 and~3.
    
    We followed the same three-rater procedure for the two framework sources added in this study. For Qiskit Algorithms, the raters classified 39 components and reached near-perfect agreement: Fleiss'~$\kappa = 0.936$ and Light's~$\kappa = 0.937$. Raters~1 and~2 agreed on all 39 components ($\kappa = 1.000$). The three disagreements involved distinctions between VQA and VQE and between Uncompute and SWAP Test. These cases were resolved through discussion.

    For Qiskit Machine Learning, the raters classified 31 candidate components and reached substantial agreement: Fleiss'~$\kappa = 0.662$ and Light's~$\kappa = 0.670$. Disagreements concerned whether components related to the Quantum Kernel Estimator should be assigned to SWAP Test or VQA and were resolved through discussion. One candidate, the former \texttt{RawFeatureVector} class, was later removed because the current public API exposes only the \texttt{raw\_feature\_vector} function. The final Qiskit Machine Learning KB therefore contains 30 components.

    For each framework source, the classifications were compared and discussed until we reached a final consensus. The framework components were classified using the 61-pattern catalog described at the beginning of this subsection. Only patterns associated with at least one framework component are represented in the KB.
 The final consensus classifications for all five framework sources are the KB itself, stored in the replication package as the CSV file \texttt{data/knowledge\_base/knowledge\_base.csv}\footnote{\url{https://github.com/saeg/qpa-v2/blob/main/data/knowledge_base/knowledge_base.csv}. The per-source classification files are in the same \texttt{data/knowledge\_base/} directory.}. Each row contains a framework source, the path of a documented API component, and its assigned pattern.  Listing~\ref{lst:kb_examples} shows one entry from each of the five framework sources.

    \begin{lstlisting}[
        style=datastyle,
        caption={Example entries from the pattern knowledge base.},
        label={lst:kb_examples}
    ]
    framework,name,pattern
    ----------------------------------------------------
    classiq,.../functions/amplitude_amplification,Amplitude Amplification
    pennylane,.../templates/embeddings/AngleEmbedding,Data Encoding
    qiskit,.../circuits/basis_change/qft/QFT,Basis Change
    qiskit-algorithms,.../amplitude_amplifiers/grover/Grover,Grover
    qiskit-machine-learning,.../neural_networks/EstimatorQNN,Quantum Neural Network (QNN)
    \end{lstlisting}

    The previous study~\citep{ramalho2026mining} built its KB from Qiskit, PennyLane, and Classiq, containing 217 components across 24 patterns. The current KB contains 286 components from five sources and represents 23 patterns. The number of components increased by 69, or approximately 32\%, through the addition of Qiskit Algorithms, with 39 components, and Qiskit Machine Learning, with 30. The number of represented patterns decreased by one after the classification audit showed that the components assigned to Schmidt Decomposition did not implement that pattern. Schmidt Decomposition remains in the 61-pattern catalog, but it is no longer linked to a component in the five framework sources.
    
    These additions do not introduce new pattern categories, since all patterns they cover already exist in the KB.  Their value is in providing new named implementations of those patterns: the tool can now recognize a practitioner importing \texttt{VQE} from \texttt{qiskit-algorithms} or \texttt{EstimatorQNN} 
    from \texttt{qiskit-machine-learning} as pattern instances, whereas the previous study would have missed these calls entirely.

\subsection{Selecting projects and notebooks}
\label{sub:github_search}
\label{sub:notebook_extraction}
To update the project population from the previous study~\citep{ramalho2026mining}, we reran and expanded the repository search on July 21, 2026. The previous search used the GitHub topics \textit{quantum-computing}, \textit{quantum-machine-learning}, and \textit{quantum-algorithms}. The expanded search also included \textit{quantum-simulation}, \textit{quantum-error-correction}, and \textit{quantum-circuit}.  The GitHub topic search does not support the logical operator ``OR'', so qpa triggers one query for each topic and consolidates the results by repository. The queries are restricted to Python-based projects only.

We selected repositories that were not forks or archived, had at least 20 stars and 10 contributors, and had been updated in the previous 12 months. Repositories serving as reading lists or collections of course materials, tutorials, and other educational resources were excluded, although eligible software projects containing notebooks or examples were kept. We also added 11 known projects to the candidate pool because topic-based searches may miss repositories that do not use the selected GitHub topics. These projects were evaluated using the same selection criteria as the other candidates. The manual candidate list is defined in the script \texttt{src/data\_acquisition/discover\_projects.py}\footnote{\url{https://github.com/saeg/qpa-v2/blob/main/src/data_acquisition/discover_projects.py}.}, while the final set of retained repositories is available in the file \texttt{data/filtered\_repo\_list.txt}\footnote{\url{https://github.com/saeg/qpa-v2/blob/main/data/filtered_repo_list.txt}.}. The filtering trace records the outcome of each selection decision.

The search found 722 candidates, excluded 639, and kept 83 repositories. The trace for this run, including the query results and the reason for every exclusion, is stored in the replication package as the text file \texttt{data/github\_search\_summary\_20260721\_111315.txt}\footnote{\url{https://github.com/saeg/qpa-v2/blob/main/data/github_search_summary_20260721_111315.txt}.}.

The 83 repositories returned by the search were adjusted before the analysis. Qrisp was reserved for the held-out evaluation, and the archived \texttt{tensorcircuit} repository was removed because its notebooks are duplicated in \texttt{tensorcircuit-ng}. Classiq Library and Qiskit Algorithms were then added as framework example sources. These two exclusions and two additions produced a complete mixed corpus of 83 repositories. For the adoption analysis, we excluded Classiq Library, Qiskit Algorithms, and Qiskit Machine Learning because they are maintained by framework providers represented in the KB. The resulting set contains 80 repositories. The complete mixed corpus is retained as a secondary artifact. 

Our analysis focuses on proxies for practical applications, which are typically demonstrated in Jupyter Notebooks rather than in the core framework code. To analyze these examples, the qpa tool performed a two-step process. First, it scanned the selected projects to find all notebooks. Second, it converted them into Python scripts so that we could analyze the code using Abstract Syntax Trees (ASTs).

The conversion process produced 1,144 Python scripts in the complete mixed corpus, up from the 985 Jupyter Notebooks analyzed in the previous study across 80 projects. Qrisp was held out for evaluation, and \texttt{tensorcircuit} was excluded because it is the archived original of \texttt{tensorcircuit-ng} and contains duplicate notebooks. The mixed corpus contains 83 repositories, of which 50 contributed scripts. 

After excluding the three framework-maintained example repositories, the resulting set contains 80 repositories, 47 of which contributed 904 scripts to the adoption analysis.

\begin{figure}[H]
    \centering
    \includegraphics[
        width=\linewidth,
        height=0.78\textheight,
        keepaspectratio
    ]{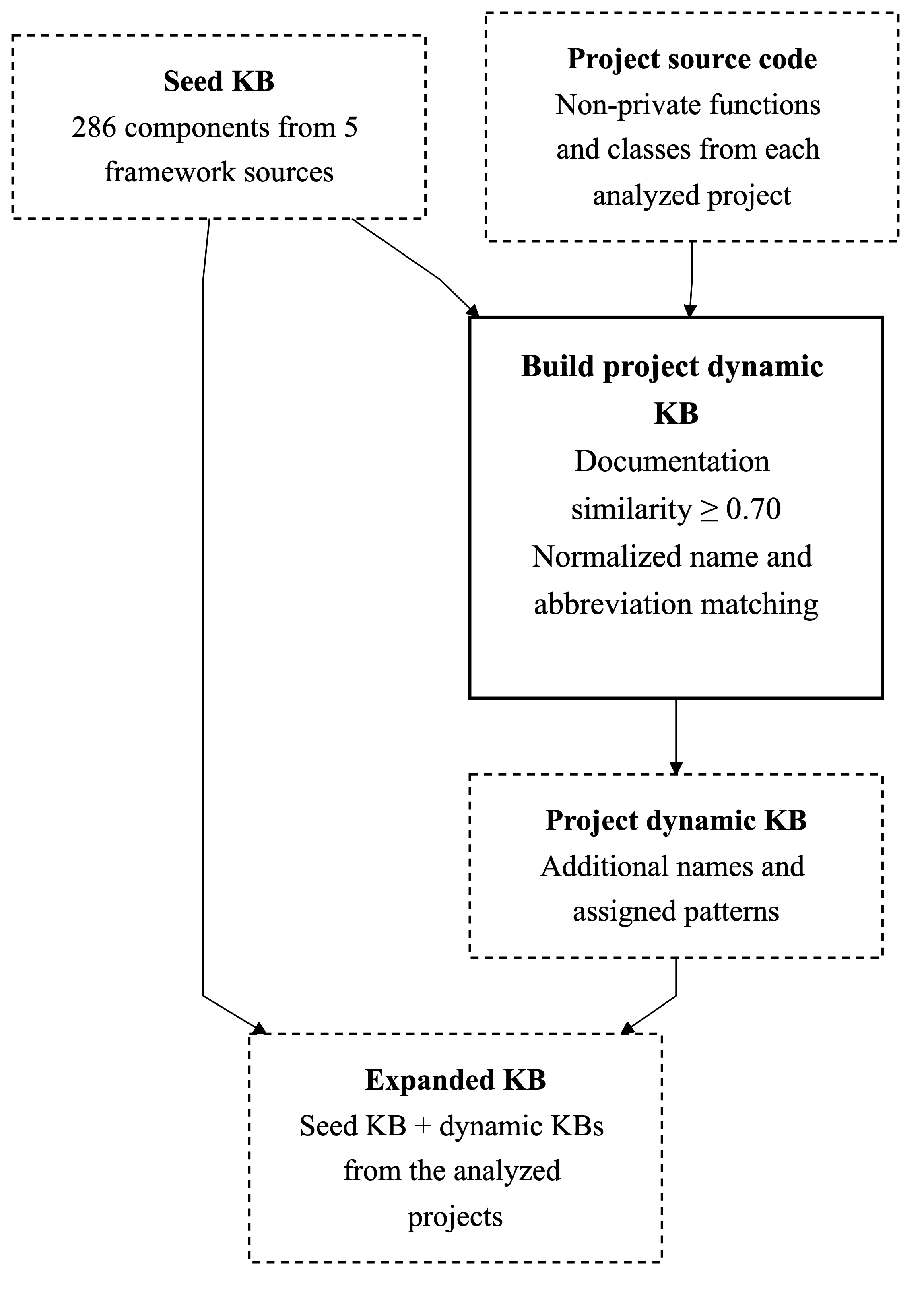}
    \caption{Vocabulary expansion. qpa builds project-specific dynamic KBs and combines them with the seed KB before notebook analysis.}
    \label{fig:diagram3a}
\end{figure}
\begin{figure}[H]
    \centering
    \includegraphics[
        width=\linewidth,
        height=0.78\textheight,
        keepaspectratio
    ]{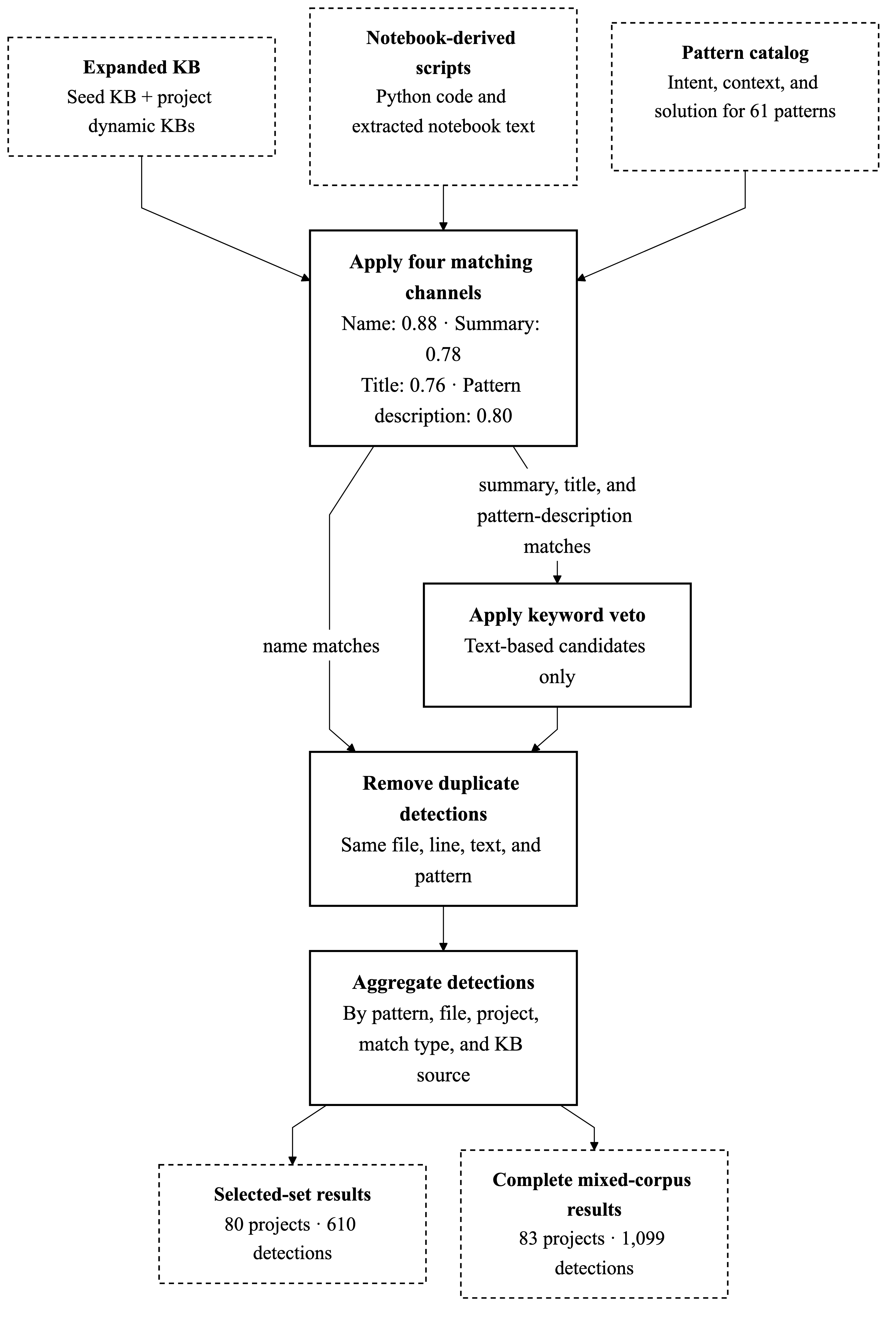}
    \caption{Pattern detection and aggregation. qpa applies four matching channels, filters text-based candidates, removes duplicate detections, and aggregates the results.}
    \label{fig:diagram3b}
\end{figure}

\subsection{Matching patterns to code}
\label{QCPatternMatchingGithub}

qpa analyzes the 904 scripts extracted from the selected set of 80 projects by comparing their contents with the framework components and pattern assignments in the KB described in Section~\ref{sub:pattern_kb}. Figures~\ref{fig:diagram3a} and~\ref{fig:diagram3b} summarize the process. qpa prepares the KB for semantic comparison, expands its vocabulary using function and class names extracted from project source code, and then applies four matching channels to the converted notebooks. Finally, it filters weak text-based matches and removes duplicate detections.

    \noindent \textbf{Preparation of the knowledge base.} This consisted of converting the framework-component records in the KB into numerical embeddings using the \textit{all-mpnet-base-v2} Sentence Transformer model \citep{allmpnetbasev2}. For each component, we generated three sets of vectors: one from its short name, one from its docstring summary, and one from the intent description of its associated pattern. The model supports up to 384 word-piece tokens per input, avoiding the silent truncation of long docstrings that shorter-context models introduce. The embedding model was chosen for replicability: it runs locally, avoiding the costs and non-deterministic outputs associated with API-based large language models \citep{LLMHallucination2025}.

    \noindent \textbf{Semantic search and matching.}
    The previous version of qpa used name and summary matching. In this study, we add title matching and pattern-description matching, resulting in four channels. All four channels use cosine similarity over embeddings produced with the \textit{all-mpnet-base-v2} model:

    \begin{enumerate}
        \item \textbf{Name matching} (threshold: 0.88): embeddings of function calls extracted via AST are compared against normalized component names.
        \item \textbf{Summary matching} (threshold: 0.78): an embedding of all comments in a file is compared against component documentation summaries.
        \item \textbf{Title matching} (threshold: 0.76): the first Markdown heading and its introductory paragraph, extracted as comments from the converted notebook, are matched against component summaries.
        \item \textbf{Pattern description matching} (threshold: 0.80): the file's comment block is compared against the intent, context, and solution descriptions of each pattern.
    \end{enumerate}

    A match is recorded when one of the channels reaches its threshold. The channel name and similarity score are saved in the output for analysis. 

    After matching, an automatically generated keyword veto is applied to candidates produced by the three text-based channels: summary, title, and pattern description. The keyword lists are generated from the knowledge base rather than selected manually. For each candidate component match, the tool combines its summary, docstring, internal keywords, and internal comments, splits this text into lowercase tokens, and removes Python terms, common English words, generic quantum-computing terms, and tokens shorter than three characters. It then keeps, for each pattern, up to 30 of the most frequent tokens that occur at least twice. A token is considered discriminating when it occurs in the retained lists of no more than two patterns. The candidate is retained only if the source file contains at least one discriminating keyword from that pattern's list. Otherwise, it is excluded. Name matches are the only channel that does not go through this process. If no keyword list is available for a pattern, the candidate is retained. The veto helps reduce false positives by checking whether a semantic match is also supported by words related to the pattern. Limiting a keyword to at most two pattern lists reduces matches based on generic words.

    \noindent \textbf{Counting unit.} The same match may be found through multiple KB entries because the seed and dynamic KBs can overlap. To remove these duplicates, qpa treats matches with the same file, line, matched text, and pattern as one detection. It keeps the match with the highest similarity score and records how many KB entries produced the same evidence. A pattern can still have more than one detection in a file when it is found on different lines or through different textual evidence. In the results, detection counts refer to these de-duplicated records, pattern frequency refers to unique (file, pattern) pairs, and project coverage is the number of projects in which a pattern was found.

    \noindent \textbf{Vocabulary expansion (two-phase pipeline).}
    The four matching channels compare the code in the selected projects with the five-source seed KB. However, projects that use other frameworks may refer to the same quantum operations using different function or class names. These differences can cause qpa to miss pattern implementations. To reduce this problem, qpa expands the KB using the source code of the selected projects before analyzing their notebooks. 
    
    For each eligible project, qpa extracts non-private functions and classes from its Python source code and tries to associate them with patterns in two ways. First, it compares their documentation with the descriptions of the components already classified in the seed KB. If the highest similarity is at least 0.70, qpa assigns the same pattern as the most similar component. Second, qpa compares each function or class name with the names of the 23 patterns represented in the seed KB. These pattern names come from the existing component-to-pattern assignments and are a subset of the 61-pattern catalog. qpa automatically splits function, class, and pattern names into words, converts them to lowercase, and removes generic terms such as \textit{quantum}, \textit{algorithm}, and \textit{circuit}. It also extracts abbreviations written in the pattern names, such as QPE and QAOA. A name is associated with a pattern when it contains the pattern's abbreviation or all of its remaining distinctive words. For example, this comparison can associate names such as \texttt{QPE}, \texttt{QAOAProblem}, and \texttt{grovers\_alg} with their respective patterns. When the two comparisons produce different assignments for the same name, qpa keeps the assignment based on the documentation.
    
    For each project, this process produces an additional list of function and class names and their corresponding patterns. We refer to this list as the project's dynamic KB. After processing all eligible projects, qpa combines their dynamic KBs with the seed KB and uses the expanded KB to analyze the notebooks.
    The final output is a CSV file\footnote{Primary adoption dataset:
    \url{https://github.com/saeg/qpa-v2/blob/main/data/third_party_quantum_concept_matches_with_patterns.csv};
    complete mixed-corpus dataset:
    \url{https://github.com/saeg/qpa-v2/blob/main/data/quantum_concept_matches_with_patterns.csv}.}
    where each record contains the file path, matching KB component, its pattern, the match type, 
    the triggering text, and the similarity score.


    The analysis generates a CSV file of pattern matches, which the qpa tool uses to produce a consolidated report covering high-level statistics, the sources of the framework components responsible for the detections, pattern prevalence across the ecosystem, and identification of patterns not found in any project.

\subsection{Language-model support for knowledge-base maintenance}
\label{sec:kb_drift}

    The framework sources used to build the KB continue to evolve, and newer versions may contain components that were not included in the manual categorization. Keeping the KB up to date would therefore require these components to be identified and assigned to patterns. qpa stores newly extracted components that do not yet have a pattern assignment in a separate file for later review. We investigated whether an ensemble of language models could assist with their classification.

    This experiment is separate from the main detection and composition analyses. The components considered here and their language-model assignments were not added to the 286-component KB used in the results reported in this paper. Instead, the experiment evaluates whether language models could support future updates to the KB.

    \noindent \textbf{Prompt and model configuration.}
       Each new component was classified independently by three instruction-tuned
       chat models, drawn from three different providers, and all served through
       the Together~AI \citep{TogetherAI} inference platform:
       \texttt{meta-llama/Llama-3.3-70B-Instruct-Turbo},
       \texttt{Qwen/Qwen2.5-7B-Instruct-Turbo}, and
       \texttt{deepseek-ai/DeepSeek-V4-Pro}. Every model received the same two-part
       prompt: a fixed \emph{system} message that set the role and configured the output
       format, and a component-specific \emph{user} message that supplied the framework component together with the catalog. With the component fields and the catalog shown as substitution slots, the prompt was:

       \begin{quote}\small\ttfamily\raggedright\noindent
       [system] You are an expert in quantum computing and quantum software design
       patterns. You classify a quantum framework's component into exactly ONE
       pattern from a fixed catalog. Answer ONLY with a single JSON object and
       nothing else.\\[4pt]
       [user] Classify the following quantum framework component into exactly one
       pattern from the catalog.\\
       COMPONENT\\
       - framework: \{framework\}\\
       - name: \{short name\}\\
       - fully-qualified name: \{fully-qualified name\}\\
       - docstring summary: \{summary\}\\
       PATTERN CATALOG (choose exactly one "pattern" -- copy the catalog name
       verbatim):\\
       \{the 61 patterns, one "- name: definition" line per pattern\}\\
       Respond with strictly this JSON and nothing else:\\
       \{"pattern": "\textless{}exact catalog name\textgreater{}", "confidence":
       \textless{}number between 0 and 1\textgreater{}, "reason":
       "\textless{}one short sentence\textgreater{}"\}
       \end{quote}

    We also checked the length of the prompts used to classify the 50 components. Each prompt contained the component's framework, name, fully qualified name, documentation summary, and the names and definitions of the 61 patterns. Using the \texttt{cl100k\_base} tokenizer, the prompts contained an average of approximately 1,913 tokens, with a minimum of 1,883 and a maximum of 1,963. These values are well below the context limits of the selected models.

       All requests used one fixed decoding configuration. The sampling temperature was set to $0$ (greedy decoding, with no randomness), the completion was capped at $400$ tokens, and the answer was requested as a JSON object (the
       API's \texttt{json\_object} response format) for every model that supports
       it. Each reply was parsed for its \texttt{pattern} field and mapped back onto
       the catalog by a case- and punctuation-insensitive comparison that also
       accepts a pattern's parenthetical abbreviation; a reply that failed to parse
       as JSON, or that named a pattern outside the catalog, was re-requested up to
       three times and otherwise discarded. Every accepted answer was cached on disk
       under a (model, component) key, so the ensemble can be resumed or re-run
       without re-querying the models and without changing the assigned labels. 

\begin{figure}[h!]
      \centering
      \includegraphics[width=0.7\textwidth]{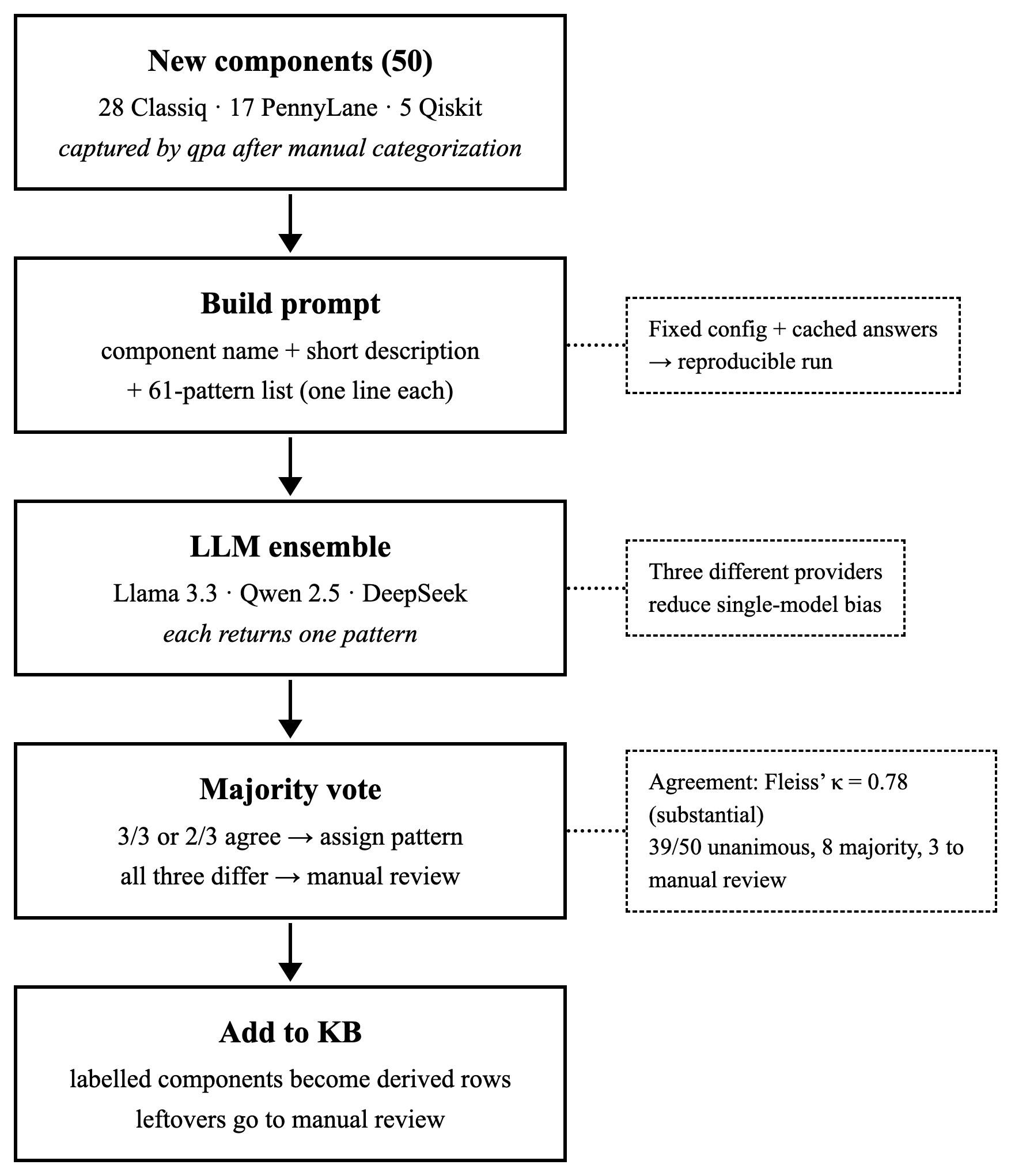}
      \caption{Automated pattern categorization with an LLM ensemble. The new components collected by the tool after the manual categorization are turned into a prompt (the component name, its documentation summary, and the 61-pattern list), classified independently by three models from different providers (Llama~3.3, Qwen~2.5, and DeepSeek), and combined by majority vote: a component on which at least two models agree receives that pattern, while a component on which all three differ is left for manual review.}
      \label{fig:llmflow}
    \end{figure}

       The three models were combined by majority vote: a component on which at
       least two models agreed received that pattern, and a component on which all
       three differed was left for manual review. We applied this to the 50 newly extracted components the tool had collected since the last manual categorization (28 from Classiq, 17 from PennyLane, and 5 from Qiskit).
       Figure~\ref{fig:llmflow} summarizes these steps.

       Using an LLM ensemble this way trades some explainability and determinism for automating a step that is otherwise entirely manual. Drawing the three models from different providers also reduces the chance that the outcome reflects a single provider's bias. Together~AI is a paid service, which is a barrier to exact replication, but because the task is a fixed assignment over a closed 61-pattern catalog, its outputs are easy to audit and reproduce by hand for readers without API access.

    The three models agreed completely on 39 of the 50 components (78\%). On 8 further components, two of the three agreed, and on the remaining 3 components each model chose a different pattern. To measure the agreement between the three models we computed Fleiss' kappa (0.78), the same measure used for the manual categorization in the previous study. The three pairwise Cohen's kappa values were 0.87, 0.77, and 0.71, with an average of 0.78. On the commonly used scale, values in this range correspond to substantial agreement.

    We assigned the three components with tied votes by author consensus, following the same discuss-to-consensus procedure used for disagreements during the original manual classification. We therefore do not compute a separate Fleiss' kappa over these three items, which would not be meaningful at that size. \texttt{MultiplexerStatePreparation} is assigned \textit{Initialization}, matching the other arbitrary state-preparation templates in the knowledge base (\texttt{MottonenStatePreparation},  \texttt{ArbitraryStatePreparation}, and \texttt{QROMStatePreparation}). \texttt{BBQRAM} and \texttt{HybridQRAM} are assigned \textit{Function Table}, matching the QRAM variants already in the knowledge base (\texttt{SelectOnlyQRAM} and \texttt{FFQRAM}): their defining operation
    $\sum_i \alpha_i\,|i\rangle|0\rangle \;\rightarrow\; \sum_i \alpha_i\,|i\rangle|b_i\rangle$
    is an address-to-data lookup, which is consistent with the Function Table pattern.

    The disagreement of the language models is in itself informative. Two of the three tied components, \texttt{BBQRAM} and \texttt{HybridQRAM}, belong to the quantum-memory family, and that family is itself split across the catalog: the QRAM read/write primitives fall under Function Table, the QROM subroutine under \textit{Data Encoding}, and
    \texttt{QROMStatePreparation} under \textit{Initialization}. A component that the single-label vote cannot classify, together with inconsistent labels assigned to related components, may indicate that the catalog lacks a dedicated pattern for addressable quantum memory, such as QRAM or QROM.  This is a quantum routine that appears in other frameworks (Classiq's lookup tables, PennyLane's QRAM and QROM templates). We do not update the 61-pattern catalog here, as it is the rated artifact
    reused from the previous study, but we record the consolidation of the QRAM/QROM family into a single pattern as a candidate for future catalog work. The disagreement between the three models helped us identify a possible gap in the pattern taxonomy.
    
    We also used the ensemble to double-check the manual classification from the previous study, running it on the original KB of 217 components and 24 patterns. The models received each component's source and documentation but not its existing label, with a fourth model breaking ties. The ensemble labeled 214 components and reproduced the existing label for 160 of them (accuracy and micro F1 of 0.748, macro F1 of 0.615, Fleiss' kappa of 0.803 among the three base models). However, Schmidt Decomposition was missing from the ensemble's output: the four components that carried this label in the original KB all received other labels.

    We therefore reviewed these four components at the source level and confirmed the mismatch: \texttt{MottonenStatePreparation} implements Initialization, \texttt{ArbitraryUnitary} implements Circuit Construction Utility, and \texttt{HilbertSchmidt} and \texttt{LocalHilbertSchmidt} implement Hilbert--Schmidt cost tests, i.e., SWAP Test. None implements the state preparation defined by the Schmidt Decomposition pattern. We corrected the four assignments and reran the KB consolidation, vocabulary expansion, detection, evaluation, and graph analyses. The resulting KB contains 286 components covering 23 of the 61 catalog patterns. The ensemble can thus flag suspicious labels, but the final correction still depends on reviewing the source code for confirmation.
    
    \subsection{Quantum patterns in practice}
    \label{sec:results}

    We applied qpa to the selected set of open-source projects described in Section~\ref{sub:github_search}. After excluding the three framework-maintained example repositories, the resulting set contains 80 repositories. Of these, 47 contributed scripts to the scan and 37 produced at least one detection.

    \subsubsection{Overall detection results}

    qpa found at least one pattern in 358 files from 37 projects. After removing duplicates, the tool recorded 611 detections. A file can contain several pieces of evidence for the same pattern, so these detections represent 500 distinct combinations of a file and a pattern. All 23 patterns represented in the current KB were detected at least once. To focus the adoption analysis on projects outside the framework providers' own examples, we excluded Classiq Library, Qiskit Algorithms, and Qiskit Machine Learning. For completeness, the replication package also includes results for the full project set, which contains these three repositories. This set has 1,100 detections across 531 files from 40 projects.
    
    The previous study reported 573 detections in 251 files. This difference should not be interpreted as an increase in pattern adoption because the project set, KB, and detection method have changed. We therefore use the earlier result only as context for the present analysis.

    Most detections came from names of functions and classes in the source code. This channel produced 553 of the 611 detections, approximately 91\%. Of the two channels added in this study, the title channel found 48 detections, while the pattern-description channel found none. The summary channel, which was already part of the previous pipeline, found the remaining 10. The average similarity was 0.989 for name detections and 0.803 for the text-based channels. These results show that API names remain the main source of evidence. Notebook titles provided some additional coverage, but matching against general pattern descriptions did not contribute any detections under the current thresholds.

    \subsubsection{Pattern prevalence across projects}

    Table~\ref{tab:pattern_freq} lists the ten patterns with the most file-level occurrences in the selected open-source projects. A pattern is counted once for each file in which it appears, even if qpa detects it several times in that file. The table also shows the number of projects in which each pattern was found. Circuit Construction Utility appears in 87 files across 25 projects. VQA is second, appearing in 63 files across 16 projects, followed by Quantum Neural Network, which appears in 57 files across 11 projects.
    
    Compared with the previous study, variational and quantum machine learning patterns are now closer to the top of the results. VQA ranks second, QAOA and VQE are also among the top ten, and Quantum Neural Network ranks third. This does not necessarily mean that the use of these patterns has increased over time. The previous study ranked patterns by individual detections, while the present table counts each pattern once per file. The KB, selected projects, and matching procedure also differ between the two studies.

    \begin{table}[h!]
    \centering
    \caption{Top 10 patterns by frequency (unique file--pattern pairs) and project coverage.}
    \label{tab:pattern_freq}
    \begin{tabular}{lcc}
    \toprule
    \textbf{Pattern} & \textbf{File--pattern pairs} & \textbf{Projects} \\
    \midrule
    Circuit Construction Utility                       & 87 & 25 \\
    Variational Quantum Algorithm (VQA)                & 63 & 16 \\
    Quantum Neural Network (QNN)                       & 57 & 11 \\
    Quantum Logical Operators                          & 46 & 11 \\
    Data Encoding                                      & 32 & 7 \\
    Quantum Approximate Optimization Algorithm (QAOA)  & 25 & 10 \\
    Quantum Arithmetic                                 & 23 & 6 \\
    SWAP Test                                          & 22 & 10 \\
    Variational Quantum Eigensolver (VQE)              & 20 & 10 \\
    Domain Specific Application                        & 17 & 5 \\
    \bottomrule
    \end{tabular}
    \end{table}
    
    Within the selected open-source projects, the detected patterns cover the three levels of abstraction described in the previous study. Circuit Construction Utility (87 files), Quantum Logical Operators (46), and Quantum Arithmetic (23) are lower-level building blocks. VQA (63), Data Encoding (32), QAOA (25), SWAP Test (22), and VQE (20) are reusable algorithmic components or complete algorithms. Quantum Neural Network (57) and Domain Specific Application (17) are higher-level applications. These results support the earlier finding that practitioners use patterns ranging from circuit construction to complete algorithms and domain-specific applications.
    
    \subsubsection{Sources of pattern detections}

    For each detection, qpa records the KB entry that produced the retained match. When the same evidence matches entries from more than one KB source, the tool keeps the entry with the highest similarity score. These counts identify the source of the matching KB entry, not the framework imported by the analyzed file. For example, a file may be associated with the Qiskit KB because its code matches a Qiskit component, even if the file does not use Qiskit itself.
    
    Of the 611 detections, 171 were linked to components in the five-source seed KB. Qiskit accounted for 54 of these detections, Qiskit Algorithms for 39, PennyLane for 36, Qiskit Machine Learning for 26, and Classiq for 16. The remaining 440 detections, approximately 72\%, were linked to entries generated during vocabulary expansion from the source code of the selected projects.

    The 440 detections from dynamic KB entries show that qpa often relies on function and class names extracted from the selected projects, rather than only on the five seed sources. Before scanning the notebooks, the tool merges the entries extracted from all projects into one combined KB. An entry extracted from one project can therefore match a notebook from another project.
    
    These source counts cannot be used to rank framework adoption because qpa does not trace each matched call back to the package that defines it. Instead, it records the KB entry with the highest similarity. The same function or class name may occur in different frameworks, so the source of the selected KB entry may differ from the framework used by the file. Measuring framework adoption would require a separate analysis of imports, dependencies, and fully qualified calls.

    \subsection{Evaluating pattern detection}
    \label{sub:evaluation}
    To answer RQ1, we evaluate qpa against a manually labeled set of Qrisp tutorial notebooks. We first assess the embedding-based pipeline and examine the effect of vocabulary expansion. We then apply an LLM ensemble to the same notebooks and compare both detection approaches using the same ground truth and metrics.

    \subsubsection{Evaluation on Qrisp}

    To assess the accuracy of the matching pipeline, we evaluated the tool against Qrisp~\citep{qrisp}, a framework deliberately excluded from the KB, so its vocabulary is disjoint from the KB the tool matches against.
    We manually labelled 36 ground-truth file-pattern pairs spanning 11 patterns across 22 Qrisp tutorial
    notebooks (of the 23 scanned), and compared them against the tool's predictions using multi-label matching: a prediction is
    counted as correct if the corresponding pattern appears anywhere in the predicted set for that file.
    We report results for the full two-phase pipeline (seed KB plus the Qrisp dynamic KB
    built from the library source code as described in Section~\ref{QCPatternMatchingGithub}).

    Qrisp was selected as the evaluation target because it contains implementations for core quantum computing algorithms as well as for machine-learning and variational algorithms. 
    Qrisp is also designed around industrial standardisation of the quantum software ecosystem.
    Its backend communication interface is intended for submission to standardisation bodies such as DIN \citep{dinWebsite}
    and CEN/CENELEC \citep{cencenelecWebsite}, and the framework has been contributed to the Eclipse Foundation so that it maintains a
    vendor-neutral approach with community-driven governance~\citep{qrispMetaModel}.
    A companion formal meta-model formalises Qrisp's high-level programming paradigms independently of Python, laying the groundwork for implementation in enterprise languages such as C\texttt{++}, Java,
    and Rust~\citep{qrispMetaModel}.
    At the API level, Qrisp adopts a different approach from the operator-centric style of the seed KB frameworks
    (Qiskit, PennyLane, Classiq) by using classical-like typed variables
    (\texttt{QuantumFloat}, \texttt{QuantumBool}, \texttt{QuantumModulus}) and an automated uncomputation
    engine that avoids manual ancilla-qubit bookkeeping, producing circuit savings of over 55\% in CNOT count for targeted stages of Shor's algorithm~\citep{qrisp}.
    This abstraction difference (algorithm-centric naming versus operator-centric KB vocabulary) makes
    Qrisp a realistic test of vocabulary generalisation: a detector that works well in Qrisp must identify
    algorithm intent from semantic context rather than pure identifier similarity.

    \noindent \textbf{Ground-truth reliability.}
    Because Qrisp tutorial notebooks implement more than one pattern in a single file, the
    annotation task is multi-label and more ambiguous than the single-label KB classification.
    Three raters independently labelled each of the 22 notebooks with the set of patterns they observed.
    Following the same methodology used for the KB validation, we computed per-pattern binary Cohen's~$\kappa$
    across 16 observed patterns and averaged to obtain Light's~$\kappa = 0.316$, indicating fair agreement.
    This is lower than the single-label KB tasks, which is expected: a multi-label setting introduces more ambiguity because raters must also decide how many patterns are present.
    The exact set match rate (all three raters assign the identical pattern set) was only 18.2\% (4/22).
    The main sources of disagreement were: (i)~the Block Encoding tutorials (\texttt{BE\_vol1}, \texttt{BE\_vol2}),
    where two raters labelled the dominant operation as Data Encoding / Matrix Encoding while the
    final label is Linear Combination of Unitaries, since the Block Encoding primitive is a specialised LCU;
    (ii)~notebooks such as \texttt{Shor.py}, \texttt{Sudoku.py}, and \texttt{TSP.py}, where one rater applied
    the Domain Specific Application catch-all pattern rather than the specific sub-patterns (Grover, QPE,
    Amplitude Amplification); and (iii)~\texttt{HHL.py}, which involves Initialization, QPE, and Hamiltonian
    Simulation, leading to different choices about which patterns to include.
    All disagreements were resolved in a final consolidation session in which the raters discussed each
    ambiguous notebook and aligned on the set of patterns present, producing the 36-pair ground truth used
    for evaluation.

    \begin{table}[h!]
    \centering
    \caption{Per-pattern precision, recall, and F1 on the Qrisp evaluation set
    (GT\,=\,36 file-pattern pairs, 11 patterns). Two-phase pipeline: seed KB augmented with
    Qrisp dynamic KB (expansion threshold 0.70). ``--" indicates an undefined metric.}
    \label{tab:evaluation}
    \footnotesize
    \begin{tabular}{lrrr}
    \toprule
    \textbf{Pattern} & \textbf{P} & \textbf{R} & \textbf{F1} \\
    \midrule
    Basis Change                    & 1.000 & 1.000 & 1.000 \\
    Quantum Amplitude Estimation    & -- & 0.000 & -- \\
    QAOA                            & 1.000 & 0.875 & 0.933 \\
    Linear Combination of Unitaries & 1.000 & 0.333 & 0.500 \\
    Quantum Phase Estimation (QPE)  & 1.000 & 0.667 & 0.800 \\
    Grover                          & 1.000 & 0.500 & 0.667 \\
    Initialization                  & 1.000 & 0.667 & 0.800 \\
    Quantum Arithmetic              & 1.000 & 0.500 & 0.667 \\
    Amplitude Amplification         & 0.000 & 0.000 & -   \\
    Hamiltonian Simulation          & 0.000 & 0.000 & -   \\
    Variational Quantum Algorithm   & -- & 0.000 & -   \\
    \midrule
    \textbf{Micro overall} & \textbf{0.913} & \textbf{0.583} & \textbf{0.712} \\
    \textbf{Macro overall} & \textbf{0.636} & \textbf{0.413} & \textbf{0.488} \\
    \bottomrule
    \end{tabular}
    \end{table}

    We report both micro-averaged and macro-averaged variants of each metric.
    \emph{Micro-averaging} pools all true positives, false positives, and false negatives across patterns before computing precision and recall, so high-frequency patterns such as QAOA (GT\,=\,8)
    and QPE (GT\,=\,6) have proportionally more influence on the aggregate score.
    \emph{Macro-averaging} computes each metric independently per pattern and then takes the unweighted
    mean across all 11 patterns, giving equal weight to rare patterns such as Quantum Amplitude Estimation
    (GT\,=\,1) regardless of how many ground-truth instances they contribute.

    As shown in Table~\ref{tab:evaluation}, the two-phase pipeline has a micro-F1 of 0.712 (P\,=\,0.913, R\,=\,0.583) and a macro-F1 of 0.488 (P\,=\,0.636, R\,=\,0.413) across 11 patterns on 36 GT pairs.
    To isolate the contribution of the vocabulary expansion, we evaluated the same pipeline with the expansion turned off, that is, the seed knowledge base alone with no dynamically discovered vocabulary, on the same 36 ground-truth pairs and with the same multi-label metric. In this configuration, the tool produces 13 detections and reaches a micro-F1 of 0.449 (P\,=\,0.846, R\,=\,0.306). Precision stays high, but recall falls from 0.583 to 0.306, because the seed names rarely match Qrisp's own vocabulary. The expansion step improves recall and raises the micro-F1 from 0.449 to 0.712. The baseline is reproduced by disabling the dynamic knowledge bases and re-running the detection on the Qrisp notebooks, as described in the replication instructions.
    Basis Change, QPE, and QAOA have clear names in the Qrisp source and all reach precision 1.000. Grover, Initialization, Quantum Arithmetic, and Linear Combination of Unitaries also have no false positives, although some ground-truth occurrences remain undetected. The two false positives are one Amplitude Amplification prediction and one Hamiltonian Simulation prediction. Quantum Amplitude Estimation and Variational Quantum Algorithm are not detected in this rerun.
    The macro-F1 of 0.488 is lower than the micro-F1 of 0.712 because each of the four patterns with no true-positive detection contributes 0.000 to the unweighted average. Overall, vocabulary expansion improves coverage for Qrisp while recall remains the main limitation.

    \begin{researchbox}
    \textbf{RQ1.} How accurately does the improved tool detect quantum patterns, and does the vocabulary expansion step help when the target framework's naming differs from the knowledge-base frameworks?

    \textbf{Answer RQ1:} On Qrisp, whose typed-variable and algorithm-centric API differs from the operator-centric vocabulary the KB was built from, the improved pipeline (the seed KB plus a dynamic vocabulary built from the target's own library source) reaches a micro-F1 of 0.712 (precision 0.913, recall 0.583), against 0.449 for the same pipeline without the vocabulary expansion step. This is the first time the tool's output is measured against a manual ground truth, since the previous study reported detection counts but never measured precision or recall. Vocabulary expansion raises recall from 0.306 to 0.583 because it learns the target framework's names from its own library before the scan. Recall remains the main limitation when the target uses naming that neither the knowledge base nor the expansion step captures.
    \end{researchbox}

    \subsubsection{An LLM ensemble as an alternative detector}
    \label{sub:llm_detector}

    The ensemble can be used as an alternative to the embedding model to detect patterns directly. We tested this on the same Qrisp set used above, under the same conditions: the same notebooks, ground truth, and scoring code. Table~\ref{tab:qrisp_llm_vs_embedding} shows
    the result. The ensemble reaches a micro-F1 of 0.704, compared with 0.712 for the embedding pipeline. These scores are close, but a set of 22 notebooks is not sufficient to claim that the approaches are equivalent. They reach similar micro-F1 values in different ways: the embedding model has higher precision but lower recall, while the ensemble finds more ground-truth patterns and also reports more false positives.

    The main difference between the two is cost. The ensemble calls a paid service
    once for every file it reads, while the embedding model runs locally at no
    extra cost. For this reason, the embedding pipeline stays the default choice,
    and the ensemble is most useful when the target framework uses names the
    embedding model does not recognize. Because the two perform at a similar level,
    little is lost by using the ensemble to extend the knowledge base, and it stays
    available as a second option for detection when needed. The models used here
    are moderate in size, so stronger models would likely raise these numbers.

\begin{table}[h]
  \centering
  \caption{Pattern detection on the Qrisp notebooks (a framework not represented in the knowledge base): the LLM ensemble versus the embedding pipeline, scored under the same task, ground truth, and metric over the 36 file-pattern pairs. The embedding pipeline has higher precision and micro F1, while the ensemble has higher recall and macro F1.}
  \label{tab:qrisp_llm_vs_embedding}
  \begin{tabular}{lcccccc}
    \toprule
    & \multicolumn{3}{c}{\textit{Micro}} & \multicolumn{3}{c}{\textit{Macro}} \\
    \cmidrule(lr){2-4}\cmidrule(lr){5-7}
    Tool & P & R & F1 & P & R & F1 \\
    \midrule
    Embedding pipeline & \textbf{0.913} & 0.583 & \textbf{0.712} & 0.636 & 0.413 & 0.488 \\
    LLM ensemble       & 0.714 & \textbf{0.694} & 0.704 & \textbf{0.638} & \textbf{0.682} & \textbf{0.632} \\
    \bottomrule
  \end{tabular}
\end{table}



\section{Pattern composition in quantum frameworks}
\label{sec:composition}

The previous section examined which patterns are detected in practitioner code. That analysis identifies the framework components referenced in a file, but it does not show how those components are implemented inside their frameworks. To examine this structure, we build a separate call graph for each of the five framework sources represented in the KB. Each graph contains the KB-classified components that can be resolved in the framework source code and the calls between them. We also connect framework-provided usage files, such as notebooks, demonstrations, and tests, to the components they call directly. Following the internal calls from these entry points shows which additional components are involved indirectly. This composition analysis is separate from the adoption counts reported in Section~\ref{sec:results} and is used to answer RQ2 and RQ3.

\subsection{Graph construction}
\label{subsec:graph_construction}

Our analysis consists of checking the KB frameworks' source code for the quantum components that were classified into patterns. The target is not only the notebooks available to demonstrate how the frameworks work, but the test suite that each framework provides. The reasoning is that in the notebooks, it could be that not every high-level component is thoroughly exercised, as notebooks typically serve more as demonstration tools, and completeness (in terms of covering every abstraction that the framework offers) is typically not a requirement. To avoid this limitation, we decided to add each framework's test suite to the analysis, since tests are more likely to exercise both high-level components and the components they call internally. 
We use Python's \texttt{ast} (abstract syntax trees) module \cite{python_ast} to inspect the source code of each high-level framework component and identify calls to other components represented in the KB. This reveals internal framework calls that are not visible in the notebooks.
For each usage file, we record the high-level framework components that it calls directly. We then inspect the source code of each component and record its calls to other KB-classified components.  Repeating this for every component produces a call graph that links a notebook to the components it uses, and each component to the sub-components it is built from. We stored this graph in Neo4j \citep{neo4j_software}, a graph database, which lets us query and draw the relationships directly.

A \texttt{CALLS} edge records a direct call: from a usage file to a
component it invokes, or from one framework component to another. A
\texttt{USES} edge records reachability: it connects a usage file to every
component that its calls eventually reach, directly or through a chain of
internal calls. For example, when a notebook calls \texttt{qpe}, the
Classiq SDK internally calls \texttt{qpe\_flexible}, then \texttt{qft},
then \texttt{qft\_no\_swap}. The notebook therefore has a single
\texttt{CALLS} edge, to \texttt{qpe}, but \texttt{USES} edges to all four
components. Since \texttt{qpe} and \texttt{qpe\_flexible} are classified
under Quantum Phase Estimation while \texttt{qft} and
\texttt{qft\_no\_swap} are classified under Basis Change, the notebook
acquires a Basis Change footprint without ever naming a Basis Change
component. This distinction between called and reached components is the
basis of the analysis in RQ2. Only components classified in the KB become graph nodes. We call the result the
\emph{projected graph} of the framework's call graph. When we built and classified the KB, we left out helper functions and purely classical glue code, so these never become nodes. When a call chain passes through a function that is not in the KB, we link the two KB components on either side with a direct edge. The projected graph is therefore not a subgraph of the framework's call graph, because such an edge stands for a path of calls and not for a single call written in the source. The measures
reported in this section (depth, role, directness ratio, and unused share) describe the projected graph and not the full public API of each framework.

\subsection{A worked example: the Classiq call graph}
\label{subsec:graph_classiq}

The Classiq package exposes 72 public functions, 64 of which are represented in the KB. Of these, 63 have source definitions that can be analyzed and are included in the composition graph. The remaining function, \texttt{suzuki\_trotter}, is a Classiq built-in whose Python source definition is not available in the analyzed package. The Classiq Library contains 216 scanned notebooks, 129 of which call at least one KB component, and the resulting Classiq's composition graph contains 53 internal edges.

\begin{figure}[!htbp]
  \centering
  \includegraphics[width=0.95\textwidth]{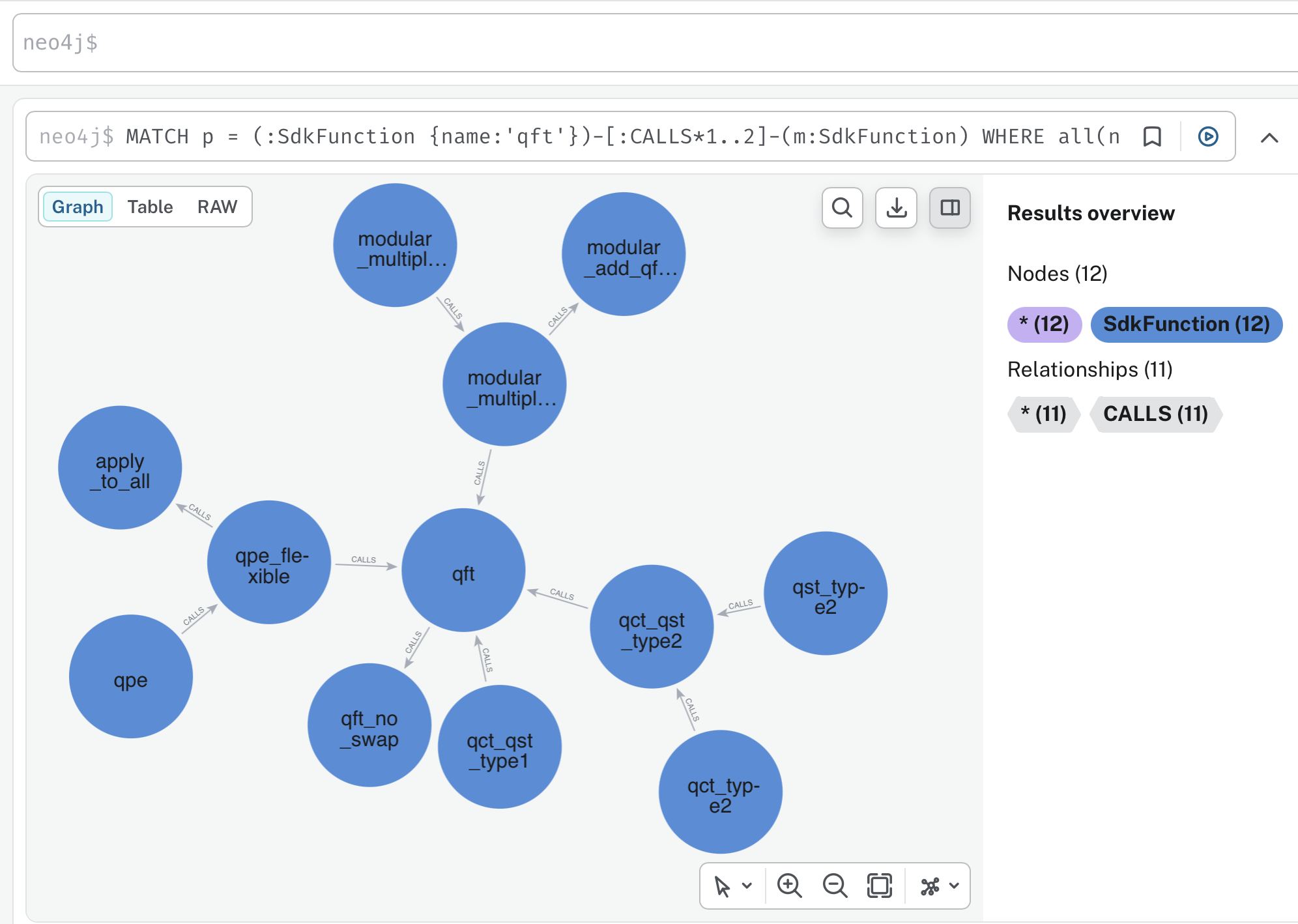}
  \caption{Framework components connected to \texttt{qft} in the Classiq SDK call graph. An arrow points from a function to a function it calls. The view includes the functions \texttt{qft} calls, the functions that call \texttt{qft}, and one further level of callers, showing how \texttt{qft} acts as a shared building block reused by higher-level routines such as phase estimation and modular arithmetic.}
  \label{fig:neo4j_qft}
\end{figure}

We use two views of the resulting graph. The first shows, for a single component, all other components connected to it. This view shows that some components serve as shared building blocks. Figure \ref{fig:neo4j_qft} does this for the Quantum Fourier Transform (\texttt{qft}): it shows both the functions that \texttt{qft} calls and the functions that call \texttt{qft}, together with one further level of callers. This shows that some components are shared primitives reused by other routines. 

\begin{figure}[!htbp]
  \centering
  \includegraphics[width=\textwidth]{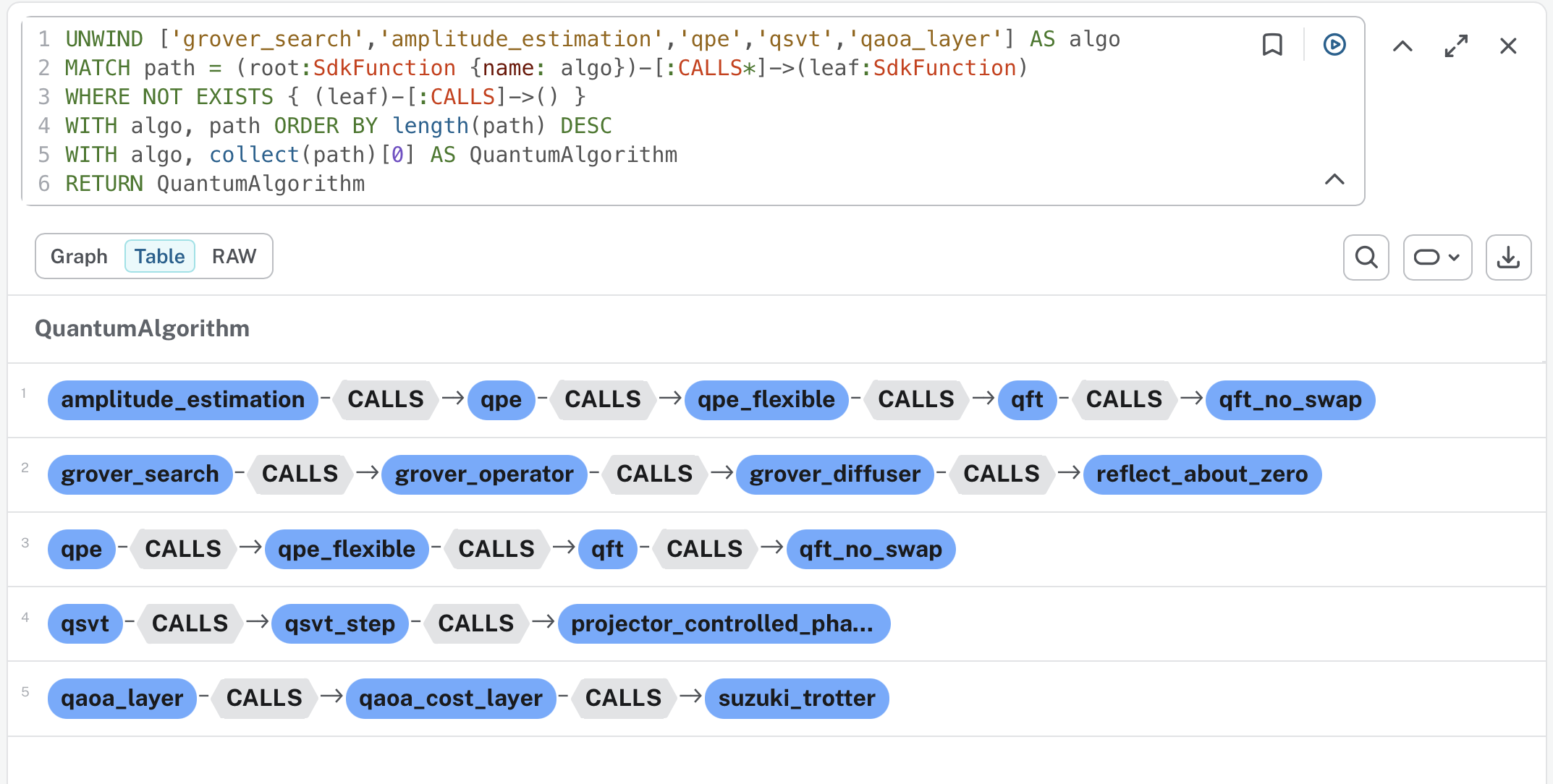}
  \caption{Decomposition of five well-known algorithms into ordered chains of sub-routines in the Classiq SDK: Grover search, amplitude estimation, quantum phase estimation, QSVT, and QAOA. Each chain follows the calls made inside the source code of a high-level function down to the lowest-level public function, showing both the composition of each algorithm and the sub-routines shared between them.}
  \label{fig:neo4j_algorithms}
\end{figure}

The second view shows how complete algorithms break down into ordered chains of sub-routines. Figure \ref{fig:neo4j_algorithms} shows this for five well-known algorithms: Grover search, amplitude estimation, quantum phase estimation, QSVT, and QAOA. For example, \texttt{grover\_search} is built from \texttt{grover\_operator}, which in turn calls \texttt{grover\_diffuser} and then \texttt{reflect\_about\_zero}. Another example, \texttt{amplitude\_estimation} reuses \texttt{qpe}, which calls \texttt{qft}. 

Storing the framework components and their call relationships in a graph database allows us to query more complex relationships. We query the graph with Cypher, Neo4j's declarative graph-query language: a query describes a small pattern of nodes joined by \texttt{CALLS} or \texttt{USES} relationships, and the database returns every part of the graph that matches it, which makes relationships like the ones below straightforward to express. For instance, Classiq's top 5 components and their respective QPs are presented in Table \ref{tab:classiq-most-referenced}. The three columns of the table are each produced by one such query (Listings~\ref{lst:classiq-usage-reach}--\ref{lst:classiq-internal-reuse}), which differ only in the relationship they count: the transitive \texttt{USES} cases, the direct \texttt{CALLS} edges coming from notebooks, and the in-degree within the SDK's own \texttt{CALLS} graph, respectively.

The column ``Usage reach (USES footprint)'' indicates the number of notebooks whose transitive footprint contains the function. The notebook does not need to call the component explicitly, as it only has to be reachable from a component that the notebook calls. The data show that \texttt{hadamard\_transform} (63) has the widest usage reach. Three of the top five components in this column carry the Basis Change pattern: \texttt{hadamard\_transform}, \texttt{qft\_no\_swap}, and \texttt{qft}. The \texttt{qft\_no\_swap} component reaches 31 notebooks but has no direct call, because it is reached through \texttt{qft}.

The column ``Direct usage (CALLS)'' shows the number of notebooks that call the function explicitly in a code cell, counting each notebook at most once. Unlike usage reach, it ignores functions that appear only transitively through another function's decomposition. \texttt{hadamard\_transform} again leads (60), and almost all of its reach is direct (60 of 63). \texttt{suzuki\_trotter} (20) appears in this column but not in the usage-reach top five; its usage reach is also 20, so every notebook that reaches it calls it directly.

The column ``Internal reuse (in-degree)'' shows the number of other SDK functions that call the component in Classiq's framework composition graph. Notebook calls are not considered. \texttt{qft} (6) has the highest internal in-degree. It is called directly in 12 notebooks but reaches 31, showing its additional role as an SDK-internal building block. This contrasts with \texttt{hadamard\_transform}, whose usage is predominantly direct.

\begin{lstlisting}[language=Cypher,
  caption={Most-referenced Classiq components by usage reach: number of notebooks
           whose transitive \texttt{USES} footprint contains each component.},
  label={lst:classiq-usage-reach}]
MATCH (u)-[:USES]->(c:ClassiqFunction)
RETURN c.name            AS component,
       c.pattern         AS pattern,
       count(DISTINCT u) AS usage_reach
ORDER BY usage_reach DESC
LIMIT 5
\end{lstlisting}

\begin{lstlisting}[language=Cypher,
  caption={Most-referenced Classiq components by direct usage: notebooks that call
           the function explicitly in a code cell.},
  label={lst:classiq-direct-usage}]
MATCH (u)-[:CALLS]->(c:ClassiqFunction)
WHERE NOT u:ClassiqFunction
RETURN c.name            AS component,
       c.pattern         AS pattern,
       count(DISTINCT u) AS direct_usage
ORDER BY direct_usage DESC
LIMIT 5
\end{lstlisting}

\begin{lstlisting}[language=Cypher,
  caption={Most-referenced Classiq components by internal reuse: in-degree in the
           SDK composition graph (how many other SDK functions call each component).},
  label={lst:classiq-internal-reuse}]
MATCH (a:ClassiqFunction)-[:CALLS]->(b:ClassiqFunction)
RETURN b.name    AS component,
       b.pattern AS pattern,
       count(a)  AS internal_reuse
ORDER BY internal_reuse DESC
LIMIT 5
\end{lstlisting}

\begin{table}[ht]
\centering
\caption{Top 5 Classiq most-referenced components.}
\label{tab:classiq-most-referenced}
\footnotesize
\renewcommand{\arraystretch}{1.3}
\begin{tabularx}{\textwidth}{@{}c X X X@{}}
\toprule
Rank & Usage reach (USES footprint) & Direct usage (CALLS) & Internal reuse (in-degree) \\
\midrule
1 & \texttt{hadamard\_transform} (63); QP: Basis Change
  & \texttt{hadamard\_transform} (60); QP: Basis Change
  & \texttt{qft} (6); QP: Basis Change \\
\midrule
2 & \texttt{apply\_to\_all} (40); QP: Circuit Construction Utility
  & \texttt{apply\_to\_all} (21); QP: Circuit Construction Utility
  & \texttt{grover\_operator} (3); QP: Amplitude Amplification \\
\midrule
3 & \texttt{qft\_no\_swap} (31); QP: Basis Change
  & \texttt{suzuki\_trotter} (20); QP: Hamiltonian Simulation
  & \texttt{hadamard\_transform} (3); QP: Basis Change \\
\midrule
4 & \texttt{qft} (31); QP: Basis Change
  & \texttt{qpe} (12); QP: Quantum Phase Estimation (QPE)
  & \texttt{qft\_no\_swap} (3); QP: Basis Change \\
\midrule
5 & \texttt{reflect\_about\_zero} (22); QP: Amplitude Amplification
  & \texttt{qft} (12); QP: Basis Change
  & \texttt{projector\_controlled\_phase} (3); QP: Phase Shift \\
\bottomrule
\end{tabularx}
\end{table}

\subsection{Analyzing pattern composition}
\label{subsec:graph_vs_detection}

The file-level detection of Section~\ref{sec:tool} answers the question of which patterns appear in a file. It counts a pattern as present whenever a component that represents that pattern can be reached from the code. This measures how patterns are used in practice, but it ignores the information preserved by the composition graphs: which component calls which and the framework in which each call occurs. We use that information first to answer the two composition research questions, and then to report two further observations about how composition affects the reading of the detection results. RQ2 asks whether a detected pattern reflects a deliberate choice by the developer or a by-product of how the framework composes its components, and we answer it through two views of the graph, direct versus indirect use and the reasons patterns co-occur. RQ3 asks at what granularity each framework exposes the same pattern. The two further observations concern patterns that become visible only at the lower levels of a high-level package, and components that are catalogued but never used.

\subsubsection{Direct and indirect pattern use (RQ2)}
\label{subsub:rq2}

We answer RQ2 through two views: separating direct from indirect use, and explaining why two patterns co-occur.

\paragraph{Direct and indirect patterns.}
\label{subsub:directness}

As described in Section  \ref{subsec:graph_classiq}, patterns can be reached directly (from notebooks) or indirectly (from internal calls). From the point of view of a practitioner, hiding lower layers is desirable, since it is the purpose of a framework to let the user write a high-level routine without managing the gate-level construction routines. The graph separates two situations that file-level detection combines: components called directly by the developer and components reached only through internal framework calls.

To make this separation measurable, we define the directness ratio. For a
component $c$, let $\mathrm{direct}(c)$ be the number of files that call $c$ directly and let $\mathrm{reach}(c)$ be the number of files whose footprint reaches $c$, whether by a direct call or through the decomposition of another component. The directness ratio of $c$ is

\begin{equation}
\label{eq:directness}
\rho(c) \;=\; \frac{\mathrm{direct}(c)}{\mathrm{reach}(c)},
\qquad \text{defined for } \mathrm{reach}(c) \ge 2 .
\end{equation}

Since every direct call also counts as reach, we have $\mathrm{direct}(c) \le \mathrm{reach}(c)$, so $\rho(c)$ lies between $0$ and $1$: a value
near $1$ means the component is almost always written by hand when it appears, and a value near $0$ means it is almost always pulled in through the implementation of something else. The restriction to components reached by at least two files keeps the ratio from being decided by a single file.

In Qiskit, several of the components with the widest reach are seldom written by hand (Table~\ref{tab:directness-hidden})\footnote{The queries and generated directness tables are available in the qpa repository at \url{https://github.com/saeg/qpa-v2/blob/main/docs/component-profile-eda.md} (section \emph{Directness ratio}).}. The component \texttt{MCGupDiag} reaches the footprint of 41 files with no direct call, and \texttt{UCRZGate} reaches 51 files with a single direct call. Both are synthesis routines that the library inserts while building other circuits. The pattern most affected is Circuit Construction Utility, which leads the Qiskit ranking through components of this kind. In Classiq, the same effect appears with \texttt{qft\_no\_swap}, which reaches 31 files without ever being called directly by the developer.

This affects how pattern counts should be compared across frameworks, because the same label can mean different things in each. In PennyLane, every component
is called directly in more than half of the files that reach it, so its pattern counts mostly reflect deliberate choices by developers. In Qiskit, much of the count for its most common pattern comes instead from components that no developer wrote explicitly, since they are inserted automatically while other circuits are being built. This shows that a raw pattern count does not, on its own, measure how often developers chose a pattern. A fair comparison of adoption across
frameworks therefore has to consider how much of each count comes from direct calls.
Table~\ref{tab:directness-direct} shows the opposite extreme: the user-facing components with $\rho = 1$ that developers always write explicitly when they appear.

\begin{table}[h!]
\centering
\caption{Components with the lowest directness ratio $\rho$ (Eq.~\ref{eq:directness})
in each framework: reached by many files but rarely called directly. Qiskit and Classiq
contain components that are pulled in purely transitively ($\rho = 0$), whereas in
PennyLane even the most indirectly used component keeps $\rho > 0.5$. Columns
``\textit{reach}'' and ``\textit{direct}'' are the number of files that reach and
that directly call the component.}
\label{tab:directness-hidden}
\footnotesize
\begin{tabular}{lllrrr}
\toprule
Framework & Component & Pattern & reach & direct & $\rho$ \\
\midrule
\multirow{3}{*}{Qiskit}
  & \texttt{MCGupDiag} & Circuit Constr. Utility & 41 & 0 & 0.00 \\
  & \texttt{UCRZGate}  & Circuit Constr. Utility & 51 & 1 & 0.02 \\
  & \texttt{UCGate}    & Circuit Constr. Utility & 42 & 3 & 0.07 \\
\midrule
\multirow{3}{*}{Classiq}
  & \texttt{qft\_no\_swap}                & Basis Change            & 31 & 0 & 0.00 \\
  & \texttt{projector\_controlled\_phase} & Phase Shift & 11 & 0 & 0.00 \\
  & \texttt{qsvt\_step}                   & Lin. Comb. of Unitaries &  9 & 0 & 0.00 \\
\midrule
\multirow{3}{*}{PennyLane}
  & \texttt{PrepSelPrep}    & Lin. Comb. of Unitaries & 11 &  6 & 0.55 \\
  & \texttt{Select}         & Dynamic Circuit         & 27 & 15 & 0.56 \\
  & \texttt{BasisEmbedding} & Data Encoding           & 38 & 22 & 0.58 \\
\bottomrule
\end{tabular}
\end{table}

\begin{table}[h!]
\centering
\caption{Components with directness ratio $\rho = 1$ in each framework: always written by hand when they appear. These are the user-facing primitives that developers select
explicitly, in contrast with the transitively reached components of
Table~\ref{tab:directness-hidden}. Columns ``\textit{reach}'' and ``\textit{direct}'' are the number of
files that reach and that directly call the component.}
\label{tab:directness-direct}
\footnotesize
\begin{tabular}{lllrrr}
\toprule
Framework & Component & Pattern & reach & direct & $\rho$ \\
\midrule
\multirow{3}{*}{Qiskit}
  & \texttt{QAOAAnsatz}     & QAOA                    & 10 & 10 & 1.00 \\
  & \texttt{LinearFunction} & Circuit Constr. Utility &  9 &  9 & 1.00 \\
  & \texttt{TwoLocal}       & VQA                     &  7 &  7 & 1.00 \\
\midrule
\multirow{3}{*}{Classiq}
  & \texttt{phase\_oracle} & Oracle                  & 11 & 11 & 1.00 \\
  & \texttt{lcu\_pauli}    & Lin. Comb. of Unitaries &  7 &  7 & 1.00 \\
  & \texttt{swap\_test}    & SWAP Test               &  6 &  6 & 1.00 \\
\midrule
\multirow{3}{*}{PennyLane}
  & \texttt{StronglyEntanglingLayers} & VQA           & 39 & 39 & 1.00 \\
  & \texttt{AngleEmbedding}           & Data Encoding & 21 & 21 & 1.00 \\
  & \texttt{GroverOperator}           & Grover        & 18 & 18 & 1.00 \\
\bottomrule
\end{tabular}
\end{table}

\paragraph{Reasons for pattern co-occurrence.}
\label{subsub:cooccurrence}

The static pattern analysis can report that two patterns appear in the same file, and counting these pairs across the dataset shows which patterns tend to be seen together (Table~\ref{tab:cooccurrence}).
Visualizing this occurrence in a graph can help us understand the reason for the co-occurrence, as the graph records which component calls which. With this information, the co-occurrence of two patterns in a file falls into one of three situations.

The first situation is when the implementation of one pattern always calls the other. In Classiq, the phase estimation routine is built on top of the quantum Fourier transform, so any file that uses phase estimation also contains the Basis Change pattern carried by the transform, whether or not the author was thinking about it, as the pairing is produced by the decomposition itself. The pairing here reflects a directional dependency between the patterns: the higher-level routine is, by its canonical construction, built from the lower one. This dependency can be definitional (for example, Grover's algorithm is a special case of amplitude amplification~\citep{Brassard_2002}, so any implementation of Grover needs to perform amplitude amplification) or just conventional, as with phase estimation and the Fourier transform.

These dependencies vary across frameworks in how the routine is
built and are not strictly necessary. For example, the standard phase estimation routine uses an inverse quantum Fourier transform, but iterative versions estimate the same phase with a single ancilla qubit and classical feedback, without any Fourier transform~\citep{Dob_ek_2007}. Both versions exist in qiskit-algorithms: \texttt{PhaseEstimation}, which uses the QFT, and \texttt{IterativePhaseEstimation}, which does not. The co-occurrence we observe therefore reflects the implementation the framework chose, not a requirement of the algorithm. For the same reason, the graph only sees this kind of dependency when the framework provides the lower routine as a separate component. When it does not, the lower routine is written inline and leaves no call edge, so the pattern is present in the code but does not appear in the graph. Cirq, for example, has no separate Grover routine, so a Grover implementation there would perform the amplitude amplification without producing an edge to it. The graph therefore cannot identify these dependencies. 

\begin{figure}[!htbp]
  \centering
  \includegraphics[width=\textwidth]{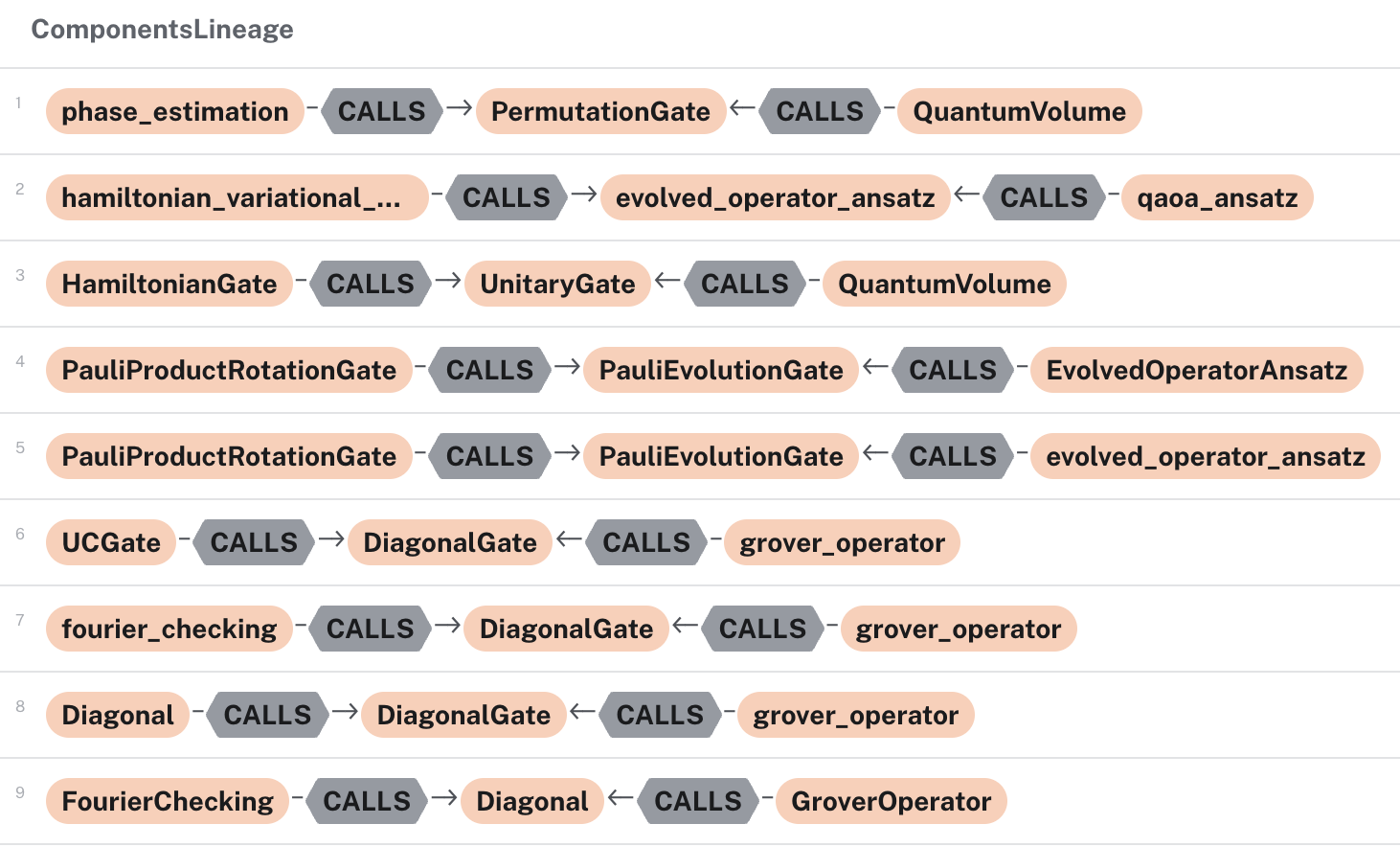}
  \caption{Shared building blocks in Qiskit. Each triple is two components of
  different patterns (the outer nodes) that both call the same
  lower-level component (the centre node); an arrow points from a function to a
  function it calls. For example, the Hamiltonian Simulation routine
  \texttt{HamiltonianGate} and the Circuit Construction Utility routine
  \texttt{QuantumVolume} both call \texttt{UnitaryGate}, a gate-building
  component. Their two patterns therefore co-occur in any file that uses either
  routine, even though neither calls the other. 
  }
  \label{fig:qiskit-cooccurrence-shared}
\end{figure}

The second situation is when two patterns are not directly related, but both are implemented as two lower-level routines. In Qiskit, Hamiltonian Simulation and Circuit Construction Utility appear together because Hamiltonian routines are built from the same basic gate-building parts that many other routines use too. The two patterns are linked through these shared parts, and not because one calls the other. 
Figure~\ref{fig:qiskit-cooccurrence-shared} shows the result, where \texttt{HamiltonianGate} (Hamiltonian Simulation) and \texttt{QuantumVolume} (Circuit Construction Utility) both call the shared component \texttt{UnitaryGate} (row 3).


The third situation is when the author combined two patterns that do not depend on each other, which is the case that the static count was implicitly assuming for every pair.

\begin{table}[h!]
\centering
\caption{Top five co-occurring pattern pairs in the usage files analyzed for Classiq, Qiskit, and PennyLane.}
\label{tab:cooccurrence}
\footnotesize
\begin{tabular}{llr}
\toprule
Framework & Pattern pair & Files \\
\midrule
\multirow{5}{*}{Classiq}
  & Basis Change \& Circuit Constr. Utility   & 36 \\
  & Basis Change \& QPE                       & 18 \\
  & Circuit Constr. Utility \& QPE            & 18 \\
  & Amplitude Amplification \& Basis Change   & 16 \\
  & Linear Combination of Unitaries \& Phase Shift & 11 \\
\midrule
\multirow{5}{*}{Qiskit}
  & Circuit Constr. Utility \& Hamiltonian Simulation & 14 \\
  & Basis Change \& Circuit Constr. Utility           &  8 \\
  & Circuit Constr. Utility \& Initialization         &  5 \\
  & Basis Change \& Hamiltonian Simulation            &  5 \\
  & Hamiltonian Simulation \& QAOA                    &  4 \\
\midrule
\multirow{5}{*}{PennyLane}
  & Data Encoding \& Dynamic Circuit                  & 23 \\
  & Basis Change \& QPE                              & 18 \\
  & Data Encoding \& VQA                             & 18 \\
  & Data Encoding \& Linear Combination of Unitaries & 17 \\
  & Basis Change \& Data Encoding                    & 16 \\
\bottomrule
\end{tabular}
\end{table}

Table~\ref{tab:cooccurrence} lists the strongest pairs in each framework, and it shows that a high count does not by itself tell us which of the three situations produced the pairing. In Qiskit, the most frequent pair joins Circuit Construction Utility and Hamiltonian Simulation. It is a shared-routine pairing, the second situation above, shown in Figure~\ref{fig:qiskit-cooccurrence-shared}. In Classiq, the most frequent pair joins Basis Change and Circuit Construction Utility, but it is of a different nature. It pairs the two most used components of the library, \texttt{hadamard\_transform} and \texttt{apply\_to\_all}, which the graph does not connect by any call. Their pairing reflects developers using both utilities in the same file rather than the framework building one from the other, so it belongs to the third situation. Reading the count together with the call structure is what tells the three situations apart. The pairings that point to deliberate composite use are those of the third situation, and they become easier to identify once the forced and shared-routine pairings have been set aside.

\begin{researchbox}
\textbf{RQ2.} Do the patterns that detection finds in a file reflect deliberate choices by the developer, or are they partly a by-product of how the framework composes its components?

\textbf{Answer RQ2:} Both, and a raw count does not reveal which. A pattern can appear in a file because the developer wrote a component of it, or because the framework pulls such a component into the implementation of something else they called. The graph separates these two origins through the directness ratio: in PennyLane, most components are written by hand when they appear, while in Qiskit much of its most common pattern comes from components inserted automatically while a circuit is built. Co-occurrence works the same way: two patterns can appear together because one is built from the other, or because both use a shared lower-level component, rather than because the developer combined them on purpose. With the graph, we can separate deliberate use from framework structure.

\end{researchbox}

\subsubsection{The granularity of pattern implementation (RQ3)}
\label{subsub:structure}

Organizing the framework components as call graphs also allows us to see how a framework's components relate to one another, i.e., whether the framework is built as a set of components that call one another and span different inner layers, or as a flat collection of independent primitives. That shape is a property of the framework itself, and the call graph can make it visible. We summarize it with two measures. \textbf{(1) The depth of a component.} It is the length, in hops, of the longest chain of internal calls that starts at it. If a component does not call another component, it is a leaf of depth zero. \textbf{(2) The component's role.} Each component is also placed in one of four roles by whether other components call it and whether it calls others: (i) a hub does both; (ii) a building block is only called; (iii) an orchestrator only calls, and (iv) an isolated component does neither. Table~\ref{tab:framework-shape} reports these across the five frameworks.



\begin{table}[h!]
\centering
\caption{Structural shape and observed usage of each framework's projected graph. Depth is the longest chain of retained internal calls, in hops ($0$ = leaf). The four structural roles partition the components
and sum to 100\% within each framework, subject to rounding. Unused is an independent measure: the share of components for which no direct or transitive call was detected in the analyzed usage files.}
\label{tab:framework-shape}
\scriptsize
\setlength{\tabcolsep}{3.5pt}
\begin{tabular}{@{}lrrr@{\hspace{6pt}}rrrr@{\hspace{6pt}}r@{}}
\toprule
& & \multicolumn{2}{c}{Depth}
& \multicolumn{4}{c}{Structural role (\%)}
& \multicolumn{1}{c}{Usage} \\
\cmidrule(lr){3-4}
\cmidrule(lr){5-8}
\cmidrule(l){9-9}
Framework & Comp. & Mean & Max
& Hub & Build. & Orch. & Isol.
& Unused \% \\
\midrule
Classiq            & 63 & 1.17 & 4 & 30.2 & 22.2 & 30.2 & 17.5 & 14.3 \\
Qiskit             & 85 & 0.88 & 7 & 11.8 & 14.1 & 20.0 & 54.1 & 11.8 \\
PennyLane          & 68 & 0.46 & 3 &  7.4 & 14.7 & 23.5 & 54.4 &  0.0 \\
Qiskit Algorithms  & 39 & 0.10 & 1 &  0.0 & 10.3 & 10.3 & 79.5 & 38.5 \\
Qiskit ML          & 30 & 0.47 & 2 & 10.0 & 13.3 & 20.0 & 56.7 & 10.0 \\
\bottomrule
\end{tabular}
\end{table}

Among the KB-classified components, Classiq is the most layered on average: its mean depth is 1.17, its longest chain has four hops, and 30.2\% of its nodes are hubs that both use and are used by others. Qiskit is wider and flatter on average: although its longest retained synthesis chain is the deepest of any framework (seven hops), 54.1\% of its KB nodes are isolated in the projected graph. PennyLane is shallower still, with a mean depth of 0.46, 54.4\% isolated nodes, and 7.4\% hubs. 

The two companion libraries look almost flat under these measures: Qiskit Algorithms has no hubs at all and a mean depth near zero. Qiskit Machine Learning is only slightly deeper. This flatness is not a sign that the packages are simple. Their components call components in the core circuit library, which is represented in a separate graph (Section~\ref{subsub:crosslayer}), so the part of their structure that would give them depth is recorded elsewhere (Qiskit core package). The measure therefore captures not only the internal shape of each framework but also the boundary at which one framework's components stop and another's begin. This is the same boundary that, as the next observation shows, hides patterns from a file-level scan.

\begin{figure}[h!]
  \centering
  \includegraphics[width=\textwidth]{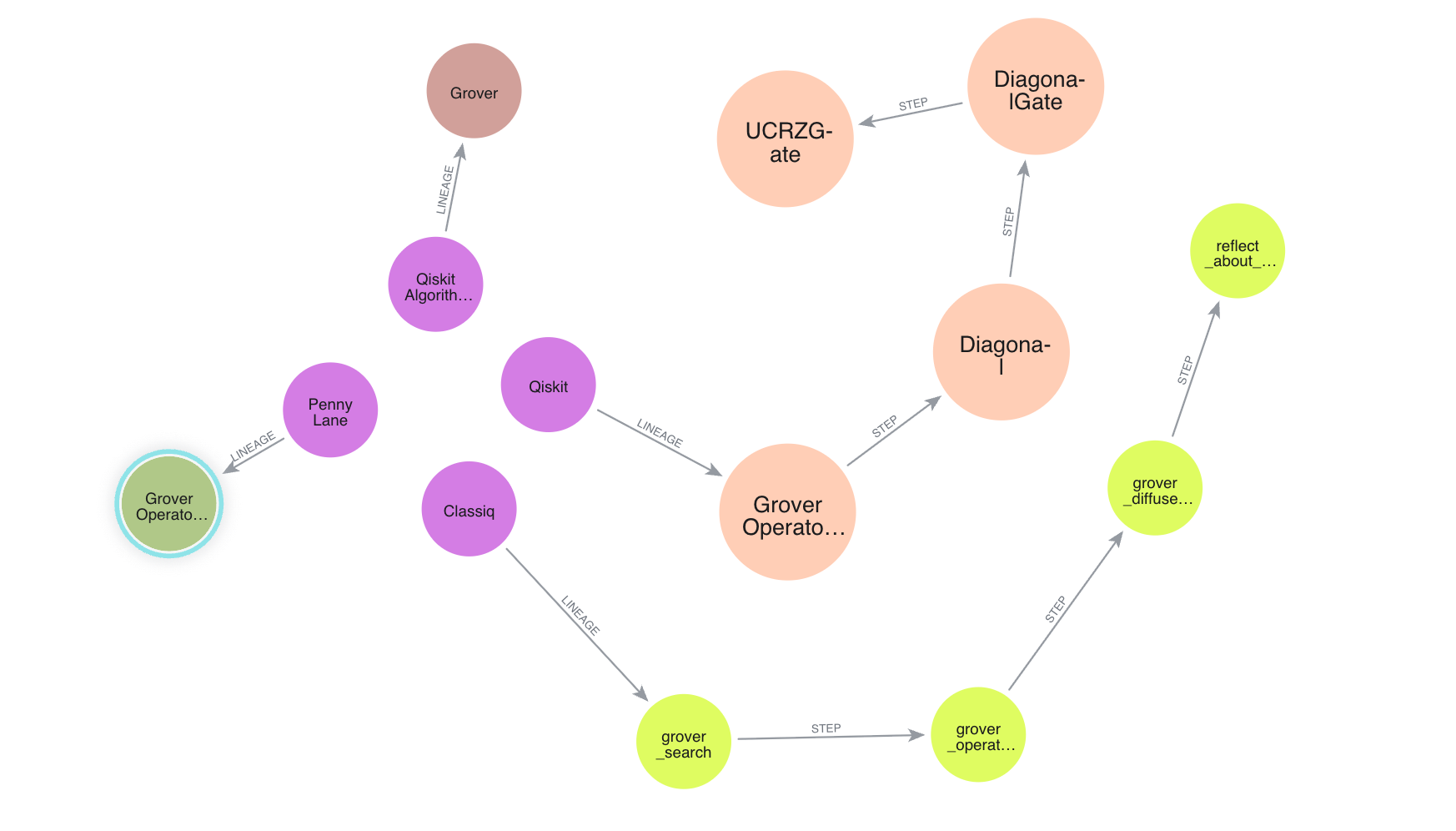}
  \caption{The Grover pattern decomposed in four frameworks, each shown as the longest chain reachable from its Grover entry point, with the framework name labelling the head of each chain. Classiq decomposes Grover into named algorithm primitives, Qiskit core into gate-synthesis components, while Qiskit Algorithms and PennyLane expose it as a single block whose internals live in a separate graph. Qiskit Machine Learning has no Grover component.}
  \label{fig:grover-lineage}
\end{figure}

Figure~\ref{fig:grover-lineage} illustrates this for the Grover pattern. It shows the longest call chain reachable from each framework's Grover entry point, with the framework name at the head of each chain. Classiq breaks Grover into a chain of named algorithm primitives, from \texttt{grover\_search} down to \texttt{reflect\_about\_zero}; Qiskit core implements its \texttt{GroverOperator} using gate-synthesis components. Qiskit Algorithms and PennyLane exposes it as a single block, since their decomposition continues in a separate graph. The same pattern therefore appears at a different granularity, and of a different kind, in each framework.

\begin{researchbox}
\textbf{RQ3.} At what level of granularity does each framework expose the same pattern, as a single high-level routine or as a chain of smaller composable components?

\textbf{Answer RQ3:} Frameworks expose the same pattern at different levels of detail. For example, Classiq represents Grover as a chain of named components (\texttt{grover\_search}, \texttt{grover\_operator}, \texttt{grover\_diffuser}, and \texttt{reflect\_about\_zero}), while Qiskit core breaks \texttt{GroverOperator} into gate-synthesis components. Qiskit Algorithms and PennyLane expose Grover as a single component whose internal structure is outside their graphs (Figure~\ref{fig:grover-lineage}). Table~\ref{tab:framework-shape} shows the difference: Classiq has the highest mean depth, while the two companion packages appear almost flat because their internal calls continue in another graph.
\end{researchbox}

\subsubsection{Further observations}
\label{subsub:further}

Beyond the two research questions, the graph supports two further observations about how composition affects the reading of detection counts.

\paragraph{Patterns that span abstraction levels.}
\label{subsub:crosslayer}

The companion frameworks Qiskit Algorithms and Qiskit Machine Learning sit one level above the core circuit library. When a file uses one of their components, the static detection records a single pattern. A call to amplitude estimation, for instance, is recorded as the Quantum Amplitude Estimation pattern. The graph shows that this single detection stands for a chain of patterns at lower levels. Amplitude estimation is implemented by calling phase estimation, and phase estimation in the core library is in turn built on the quantum Fourier transform (Basis Change pattern). One detection at the top can then be linked to two further patterns at lower levels.

This has an effect on the cross-framework comparison. The graphs for the algorithm-level packages contain no components assigned to patterns such as Basis Change and Circuit Construction Utility. This does not mean that these patterns do not exist in the workflows. Instead, they appear at a lower level of the implementation that the package-specific graphs do not cover. Without this caveat, the results could suggest that the algorithm packages avoid low-level patterns, when they actually rely on them through components provided by the core framework (Qiskit Machine Learning calling components from Qiskit, for example).

The general point is that pattern presence depends on the level at which the code is observed. A file-based scan sees the patterns named at the surface of the code. Following the call graph across the boundary between a high-level package and the core library recovers the patterns that those surface names depend on, and shows that the same workflow can be described by different patterns according to how deep the analysis looks.

\paragraph{Components that are never used.}
\label{subsub:neverused}

Static detection reports only what it finds. It can list the patterns and components that appear in the notebooks, but it says nothing about catalogued components that were never used because those leave no trace in the files. The graph includes every KB-classified component that can be resolved in the pinned framework source, including those that no usage file reaches, so it can measure this absence. It does not measure unused components outside the KB.

Across the five frameworks, most resolved KB components are reached by at least one file. In PennyLane, every resolved KB component is reached, so the catalogued pattern components and observed usage agree completely. The raw share that no file calls by name is higher in the other frameworks (Table~\ref{tab:framework-shape}), but most of those components are still present in code in a form the scan does not count as a call, so they are not necessarily a real gap. We discuss why in the threats to validity (Section~\ref{sec:threats}). A smaller set is never written anywhere in the example code, not even by name. The clearest case is Classiq, where nine concrete KB functions are absent from every notebook we scan: the register conversion routines \texttt{one\_hot\_to\_binary}, \texttt{unary\_to\_binary} and \texttt{unary\_to\_one\_hot}, the modular arithmetic routines \texttt{modular\_exponentiate} and \texttt{modular\_increment}, and four state preparation routines such as \texttt{prepare\_int} and \texttt{prepare\_sparse\_amplitudes}. These nine point to a gap between the catalogued Classiq API and its example notebooks.
All nine are nonetheless implemented in the SDK, and what is missing is their use in the example notebooks, not the implementation. This fits their nature as low-level building blocks (register conversions, modular-arithmetic primitives, and state-preparation helpers) that are typically called inside larger routines rather than demonstrated on their own. They may also be exercised by Classiq's own test suite, which we do not have available (the Classiq SDK codebase is not available on GitHub, and we downloaded it from the PyPI package manager, and that does not include test suites), or simply reflect example coverage that has not yet caught up with the API, in which case future notebooks could use them.

This view is useful in two ways. It tells the maintainer of the KB which cataloged framework components have no observed usage, which may indicate either a gap in the example code or a component that practitioners do not yet use. It also sets a limit on what a detection-based study can claim about adoption, since a component can be present in a framework and still be absent from practice, and only a view that contains the unused components can reveal that gap.

\section{Threats to Validity}
\label{sec:usage}
\label{sec:threats}

In this section, we discuss the threats to the validity of our study and the limitations of our current approach. We categorize these threats into internal, external, and conclusion validities, and list potential mitigation strategies or directions for future work.

\textbf{Internal validity.}
Regarding internal validity, a potential threat is that the pattern categorization is done by hand, which can introduce human bias. We reduced this risk by having three authors classify each framework component on their own and then discuss the cases where they disagreed until they reached a consensus. The agreement between the raters was high, with a Fleiss' Kappa of 0.8171 for the three original frameworks, and 0.936 and 0.662 for Qiskit Algorithms and Qiskit Machine Learning. A high agreement shows that the raters were consistent with each other, but it does not show that the assignment is correct, since raters who share the same background can also share the same bias. A way to reduce this dependence on manual work is to automate the categorization. The process described in Section~\ref{sec:kb_drift}, which uses an ensemble of language models from different providers to classify new components, can be extended to categorize every element of the knowledge base, and the use of models from different providers helps avoid results that depend on a single source. We used this approach as a check on the original 217-component KB. The ensemble produced final labels for 214 components and reproduced 74.8\% of the existing labels. However, this level of agreement does not prove that either classification is correct. We mitigate this with majority voting, manual review of unresolved components, and by saving every model answer to disk, so the labels we release are fixed and can be checked without calling the models again.
    
\textbf{External validity.}
The analysis focuses on Python-based projects. Thus, a possible research direction is expanding the qpa tool to support other quantum programming languages such as Q\# and Julia. In terms of external validity, a limitation of our work is the use of Jupyter notebooks to represent quantum applications. Notebooks are not large, production-level solutions, but in our work they serve as a practical proxy in the current quantum ecosystem as they show how developers use framework components to solve certain problems. With this strategy, we could identify patterns in application-level code rather than in the internal code of the frameworks themselves. As the field of quantum software engineering matures, further studies will be necessary to investigate more complex, standalone applications. Extending this analysis to other programming languages and larger software systems remains a direction for future work.

\textbf{Conclusion validity.}
Concerning conclusion validity, in terms of recall, the main limitation revealed by the evaluation (Section~\ref{sub:evaluation}) is vocabulary coverage: the KB is derived from five specific frameworks, and when a target project uses a different naming convention, the semantic channels cannot match call sites to knowledge-base components. Extending the KB with components from the target framework, or with framework-agnostic descriptions from resources such as the Quantum Algorithm Zoo \citep{QuantumZoo}, is a possible path to improving recall on out-of-distribution projects.
The same limitation explains the unused component counts of Section~\ref{subsub:neverused}. A component counts as used when a file calls it by its catalogued name, or reaches it through the internal calls of something the file does call. This is not the same as a component that no test ever runs. A test can exercise a component without naming it, for example by building a subclass of it or by calling it through a helper, and a static reader does not record that as a call. As a result, many components with a zero count are in fact present in the test code: the base class \texttt{BlueprintCircuit} appears in 117 Qiskit test files and still has a zero count, and most of the zero count components in Qiskit Algorithms are base classes and interfaces such as \texttt{MinimumEigensolver} that the tests reach through their concrete implementations. The raw zero count is therefore a lower bound on what is exercised, not a measure of dead code, and only the components that are absent even by name (the nine Classiq functions noted above) point to a real gap.

A second threat is dataset composition. In the complete mixed corpus, the Classiq Library, Qiskit Algorithms, and Qiskit Machine Learning repositories account for 489 of the 1,100 detections, or approximately 44\%. As framework-maintained examples, their matches against a KB built partly from the same frameworks are potentially self-referential. We therefore exclude all three repositories from the selected set used for the adoption analysis, which contains 611 detections across 358 files and 500 file--pattern pairs (Section~\ref{sec:results}), while retaining the complete mixed corpus as a secondary artifact. We also exclude the archived \texttt{tensorcircuit} repository because its notebooks are duplicated in \texttt{tensorcircuit-ng}. However, projects with more notebooks still have a greater influence on our adoption results.

\section{Conclusion}
\label{sec:conclusion}
The study of quantum computing patterns is an active research field, as both practitioners and academics look for ways to raise the level of abstraction beyond circuit-based programming to develop more practical applications. 
    
In this work, we presented an empirical view of how quantum software patterns are used in practice. Mining a selected set of 80 open-source projects after excluding three framework-maintained example repositories, we found 611 unique pattern detections across 358 files, showing that all 23 catalog patterns represented in our KB appear in practice, with Circuit Construction Utility and VQA retaining the broadest adoption, and that developers adopt patterns at three distinct levels of abstraction. These findings corroborate the previous adoption results while avoiding detections from framework-maintained example repositories in the selected set. Beyond detecting patterns, we constructed composition graphs from calls among framework components. We used these graphs to analyze how frameworks assemble the components associated with each pattern. The composition graphs show that raw counts mix direct developer calls with framework-introduced components and that frameworks expose the same catalogued patterns at different levels of detail.

These analyses are supported by the qpa tool and two public datasets: a KB that maps 286 framework components from five sources to the 61-pattern catalog introduced in the previous study, and a dataset describing pattern usage in open-source projects. Newly extracted components are classified by a three-model LLM vote and only enter the KB after review, so the KB can grow with the frameworks without repeating the full manual categorization. These artifacts are open-source and reusable, and are available in the qpa replication package\footnote{qpa replication package: \url{https://github.com/saeg/qpa-v2}.}. The catalog is stored as a JSON file and can be extended with additional patterns or anti-patterns, while the pipeline can be applied to other Python projects. The composition graphs can also support comparisons among frameworks and help identify recurring component chains that may represent candidates for new patterns.

Future work can extend qpa to other programming languages and versions of quantum frameworks. Following calls across package boundaries could show more clearly how high-level libraries use components from core frameworks. The LLM-based classifier could be tested and improved with a larger manually labeled dataset. Analyzing calls across files and modules could also help identify architectural patterns. Finally, comparing different framework versions could show how pattern implementations and usage change over time. These extensions would allow qpa to support a broader range of studies on quantum software patterns.
\bibliography{references}

\end{document}